\documentclass{ws-ijgmmp}

\begin{document}

\markboth{Zi-Hua Weng}
{Spin Angular Momentum in Complex Octonion Spaces}

%
\catchline{}{}{}{}{}
%

\title{Spin Angular Momentum of Proton Spin Puzzle in Complex Octonion Spaces
}

\author{Zi-Hua Weng
}

\address{School of Physics and Mechanical \& Electrical Engineering
\\
Xiamen University, Xiamen 361005, China
\\
\email{xmuwzh@xmu.edu.cn
}
}



\maketitle

\begin{history}
\received{(Day Month Year)}
\revised{(Day Month Year)}
\end{history}

\begin{abstract}
  The paper focuses on considering some special precessional motions as the spin motions, separating the octonion angular momentum of a proton into six components, elucidating the proton angular momentum in the proton spin puzzle, especially the proton spin, decomposition, quarks and gluons, and polarization and so forth. J. C. Maxwell was the first to use the quaternions to study the electromagnetic fields. Subsequently the complex octonions are utilized to depict the electromagnetic field, gravitational field, and quantum mechanics and so forth. In the complex octonion space, the precessional equilibrium equation infers the angular velocity of precession. The external electromagnetic strength may induce a new precessional motion, generating a new term of angular momentum, even if the orbital angular momentum is zero. This new term of angular momentum can be regarded as the spin angular momentum, and its angular velocity of precession is different from the angular velocity of revolution. The study reveals that the angular momentum of the proton must be separated into more components than ever before. In the proton spin puzzle, the orbital angular momentum and magnetic dipole moment are independent of each other, and they should be measured and calculated respectively.
\end{abstract}


\section{\label{sec:level1}Introduction}

Up till now, there are still several problems remaining to be solved for the proton spin puzzle. In the final analysis, the proton spin puzzle is the problem about the essence of spin angular momentum. In the Quantum Mechanics, the origination and essence of the spin angular momentum remain a mystery. Recently some viewpoints are in the violent competition. The first viewpoint deems that the spin angular momentum is intrinsic, and is a quantum property without detailed classical analogy. Of course, not everyone accepts the idea. The second viewpoint argues that the spin angular momentum is not intrinsic, and is merely the property of wave-field \cite{belinfante,ohanian}, which is independent of the internal structure of particle. Furthermore, there are some other viewpoints. Undeniably these different viewpoints violate the physical model of spin angular momentum. And these conundrums have been intriguing and fazing many scholars for a long time. Only until recent years, the emergence of the electromagnetic and gravitational theories, described with the complex octonion, enables the paper to possess the possibility to clarify a part of these problems. As one of inferences, applying the complex octonion to depict the quantum theory is capable of studying the Dirac wave equation \cite{lienert}, Schr\"{o}dinger wave equation, and proton spin decomposition \cite{cheng} and so forth.

In 1843 W. R. Hamilton invented the algebra of quaternions. J. T. Graves and A. Cayley discovered the octonions independently. The quaternion and octonion will be called as the complex quaternion and complex octonion respectively, in case a portion of coordinate values are imaginary numbers, and even complex numbers. J. C. Maxwell was the first to apply the algebra of quaternions to explore the physics property of electromagnetic field. Nowadays the complex quaternion is able to depict the electromagnetic theory or gravitational theory. And the complex octonion is capable of researching the properties of electromagnetic field \cite{chanyal1} and gravitational field \cite{demir2}, including the quantum properties and curved spaces.

In recent years some scholars introduce the real and even complex quaternions to research the physics property of Quantum Mechanics. Maia \emph{et al.} \cite{maia} claimed that the Riemannian curvature of the 3-dimensional hypersurface in space-time can be represented by a quaternion quantum operator, providing a justification for quaternion quantum gravity at the Tev energy scale. Yefremov \cite{yefremov} studied the pre-geometric structure of quantum and classical particles in terms of quaternion spinors. Jiang \emph{et al.} \cite{jiang} studied the problem of diagonalization of quaternion matrices in the Quaternionic Quantum Mechanics, by means of complex representation and real representation of quaternion matrices. Marchiafava \emph{et al.} \cite{marchiafava} considered the notion of quantum quaternions as a one-parametric quantum deformation of the quaternion algebra. Mizrahi \emph{et al.} \cite{mizrahi} applied the quaternions to depict a quantum state, describing a beam of particles characterized by discrete degrees of freedom, internal parity, and spin.

Further a few scholars make use of the octonions to analyze the physics property of certain wave functions and wave equations. Edmonds \cite{edmonds} applied the three conjugations in 10-space to examine details of the operators, metric, and wave functions, in an octonion generalization of quantum electrodynamics. Benkart \emph{et al.} constructed a quantum analogue of the split octonions \cite{benkart}, by means of the natural irreducible 8-dimensional representation and the two spin representations of the quantum group. Demir \emph{et al.} \cite{demir} discussed the close analogy between electromagnetic theory and linear gravity, by the hyperbolic octonion formalism. Meanwhile Chanyal \cite{chanyal2} made an attempt to write the extension of Pauli and Dirac matrices in terms of division algebra (octonion). Furui \cite{furui} expressed a Dirac fermion as a 4-component spinor, which is a combination of two quaternions or an octonion. Bernevig \emph{et al.} \cite{bernevig} constructed one generalization of the quantum Hall effect, where particles move in an eight dimensional space under an SO(8) gauge field, by means of the algebra of octonion. Grabowski \emph{et al.} \cite{grabowski} studied a class of quantum chains, possessing a family of local conserved charges with a Catalan tree pattern, by means of the octonion. Schray \emph{et al.} \cite{schray} extended the theory of representations of Clifford algebras to employ the algebra of octonions. De Leo \emph{et al.} \cite{deleo1} applied the complex geometry to obtain a consistent formulation of octonionic quantum mechanics. And the authors \cite{deleo2,deleo3} introduced left-right barred operators to find the translation rules between octonionic numbers and $8\times8$ real matrices.

In the quantum theories describing the proton spin puzzle, at present there are four kinds of approaches to decompose the total angular momentum of proton, including the Jaffe-Manohar decomposition, Ji decomposition, Chen decomposition, and Wakamatsu decomposition. Jaffe \emph{et al.} \cite{jaffe} examined the conserved currents associated with Lorentz transformations, deriving sum rules, showing that the anomalous current is not identified with the gluon spin. Ji \cite{ji} introduced a gauge-invariant decomposition of the nucleon spin into quark helicity, quark orbital, and gluon contributions. Chen \emph{et al.} obtained the gauge-invariant spin and orbital angular momentum operators of quarks and gluons \cite{chen}, justifying the traditional use of the canonical, gauge-dependent angular momentum operators of photons and electrons in the multipole-radiation analysis and labeling of atomic states. Wakamatsu investigated the relation among known decompositions of the nucleon spin, clarifying in what respect they are common and in what respect they are different essentially \cite{wakamatsu}. Leader \emph{et al.} provided a pedagogical introduction to the question of angular momentum decomposition in a gauge theory \cite{lorce}, presenting the main relevant decompositions, discussing in detail several aspects of the controversies regarding the question of gauge invariance, frame dependence, uniqueness and measurability.

Making a carefully comparison and analysis of preceding studies, it is able to discover a few problems in the Quantum Mechanics.

1) Spin predicament. The Quantum Mechanics claims that the spin angular momentum is intrinsic. But this buck-passing theoretical explanation is devoid of legible physical images, making it hard to understand. More seriously, the Quantum Mechanics is capable of neither deciphering the origination of spin angular momentum until now, nor figuring out the proton spin puzzle. These insufficiencies urge others to propose some new theoretical explanations, attempting to uncover the essence of spin angular momentum.

2) Dubious invariants. The Quantum Mechanics affirms that the vector sum of the orbital angular momentum and spin angular momentum is an invariant. But each of four decompositions relevant to the invariant is not yet able to solve the proton spin puzzle. Meanwhile the Quantum Mechanics asserts that there is a proportional constant between the orbital angular momentum and magnetic dipole moment, excluding irrelevantly the electric/magnetic dipole moment from the scope of the invariant. The proton spin puzzle may imply that there should be a few new invariants in the proton.

3) Other parameters. The proton spin puzzle compels scholars to contemplate whether correlative studies miss several major contributions of some arguments. In other words, it may not be enough to take into account the contributions of existing influence factors. Within the proton, there may be some crucial and new constituents of the angular momentum, and even several new physical quantities. And the existing physical theories may leave out the contributions of some unknown terms or new physical quantities.

Presenting a striking contrast to the above is that it is able to utilize the complex octonion to study the quantum theory in the electromagnetic and gravitational fields, resolving a few problems derived from the proton spin puzzle.

1) Spin angular momentum. In the complex octonion space, the angular velocity of revolution can be derived from the force equilibrium equation, while the angular velocity of precession can be deduced from the precessional equilibrium equation (in Section 4). Even if the angular velocity of revolution is zero, it is still able to induce the angular velocity of precession in the electromagnetic field, generating a new component of the orbital angular momentum. The component of orbital angular momentum, associated with the angular velocity of precession, can be regarded as the spin angular momentum. Especially, one of precessional angular velocities is only one half of the angular velocity of revolution, in the electromagnetic fields.

2) Magnetic moment and angular momentum. In the complex octonion space, the total angular momentum is not suitable to become an invariant. According to the definition of octonion angular momentum, the orbital angular momentum, spin angular momentum, electric and magnetic dipole moments, and others are capable of combining together to become an invariant. Moreover, the proportional coefficient of the orbital angular momentum and magnetic dipole moment is not a constant. In other words, these two terms should be explored and measured independently. As a result, the angular momentum must be separated into more components than ever before, in the proton spin puzzle.

3) Other contributions. According to the definition of octonion angular momentum, the octonion angular momentum consists of the orbital angular momentum, spin angular momentum, electric dipole moment, magnetic dipole moment, and a few new terms. The orbital angular momentum contains not only the conventional orbital angular momentum in the classical theory, but also some new components. Meanwhile the definitions of electric dipole moment and magnetic dipole moment are extended also. Moreover, as a new influence factor, the integrating function of field potential will play partially the role of radius vector in the angular momentum. These new components will make some useful contributions to the proton spin.

In the complex octonion space, the quaternion operator and octonion field strength can be combined together to become a composite operator \cite{weng1}. From this composite operator and octonion physical quantities, it is capable of defining the octonion field source, linear momentum, angular momentum, torque, and force and so on, for the electromagnetic and gravitational fields. Subsequently it is able to combine the quaternion operator and octonion torque together to create another composite operator. From this new operator and octonion physical quantities, it is capable of exploring the physical model of spin angular momentum, Dirac wave equation \cite{penney}, Schr\"{o}dinger wave equation, and proton spin decomposition \cite{ma} and so forth.

The paper will aim to study three major constituents. a) The octonion can be separated into two parts, real part and vector part, and written as the exponential form. From the octonion exponent form, it is able to define the octonion wave function. Under a certain approximate conditions, the octonion wave function can be degenerated into the wave function in the Quantum Mechanics. b) By means of the composite operator relevant to the octonion torque in the complex octonion space, it is capable of defining the complex octonion wave function, deducing the complex octonion wave equation. The latter can be reduced into Dirac and Schr\"{o}dinger wave equations. c) From the precessional equilibrium equation, one can describe the origination, quantization, precession and attenuation of the spin angular momentum in the complex octonion space. The physical properties of spin angular momentum can be utilized to decompose the angular momentum of proton, exploring the invariant of angular momentum.

\begin{table}[h]
\caption{The multiplication of the operator and octonion quantity in the octonion space.}
\center
\begin{tabular}{@{}ll@{}}
\hline\hline
Definition                           &  Expression                                                                                                              \\
\hline
$\nabla \cdot \textbf{a}$            &  $-(\partial_1 a_1 + \partial_2 a_2 + \partial_3 a_3)$                                                                   \\
$\nabla \times \textbf{a}$           &  $\emph{\textbf{i}}_1 ( \partial_2 a_3  - \partial_3 a_2 ) + \emph{\textbf{i}}_2 ( \partial_3 a_1 - \partial_1 a_3 )
                                           + \emph{\textbf{i}}_3 ( \partial_1 a_2 - \partial_2 a_1 )$                                                           \\
$\nabla a_0$                         &  $\emph{\textbf{i}}_1 \partial_1 a_0 + \emph{\textbf{i}}_2 \partial_2 a_0 + \emph{\textbf{i}}_3 \partial_3 a_0  $        \\
$\partial_0 \textbf{a}$              &  $\emph{\textbf{i}}_1 \partial_0 a_1 + \emph{\textbf{i}}_2 \partial_0 a_2 + \emph{\textbf{i}}_3 \partial_0 a_3 $         \\
$\nabla \cdot \textbf{A}$            &  $- \emph{\textbf{I}}_0 (\partial_1 A_1 + \partial_2 A_2 + \partial_3 A_3)  $                                            \\
$\nabla \times \textbf{A}$           &  $- \emph{\textbf{I}}_1 ( \partial_2 A_3 - \partial_3 A_2 ) - \emph{\textbf{I}}_2 ( \partial_3 A_1 - \partial_1 A_3 )
                                           - \emph{\textbf{I}}_3 ( \partial_1 A_2 - \partial_2 A_1 )$                                                           \\
$\nabla \circ \textbf{A}_0$          &  $\emph{\textbf{I}}_1 \partial_1 A_0 + \emph{\textbf{I}}_2 \partial_2 A_0  + \emph{\textbf{I}}_3 \partial_3 A_0  $       \\
$\partial_0 \textbf{A}$              &  $\emph{\textbf{I}}_1 \partial_0 A_1 + \emph{\textbf{I}}_2 \partial_0 A_2 + \emph{\textbf{I}}_3 \partial_0 A_3 $         \\
\hline\hline
\end{tabular}
\end{table}

\section{Field source}

The complex quaternion space can be applied to describe the physics property of electromagnetic field or gravitational field. And the complex quaternion space for the gravitational field is independent of the complex $S$-quaternion (short for the second quaternion) space for the electromagnetic field. Further these two orthogonal complex quaternion spaces can constitute one complex octonion space. It is found that the complex octonion space is propitious to depict the physics properties of electromagnetic and gravitational fields (see Ref.[28]).

\subsection{Complex octonion space}

In the complex quaternion space $\mathbb{H}_g$ for the gravitational field, the coordinates are, $i r_0$ and $r_k$ , the basis vector is $\emph{\textbf{i}}_j$ , the complex quaternion radius vector is, $\mathbb{R}_g = i \textbf{\emph{i}}_0 r_0 + \Sigma \emph{\textbf{i}}_k r_k$. The term $\textbf{\emph{i}}_0 r_0$ is the scalar (relevant to $\textbf{\emph{i}}_0$), and the term $\Sigma \textbf{\emph{i}}_k r_k$ is the vector (relevant to $ \textbf{\emph{i}}_k$). Similarly, in the complex $S$-quaternion space $\mathbb{H}_e$ for the electromagnetic field, the coordinates are, $i R_0$ and $R_k$ , the basis vector is $\textbf{\emph{I}}_j$ , the complex $S$-quaternion radius vector is, $\mathbb{R}_e = i \textbf{\emph{I}}_0 R_0 + \Sigma \textbf{\emph{I}}_k R_k$. The term $\textbf{\emph{I}}_0 R_0$ is the scalar-like quantity (relevant to $\textbf{\emph{I}}_0$), and the term $\Sigma \textbf{\emph{I}}_k R_k$ is the vector-like quantity (relevant to $\textbf{\emph{I}}_k$). Subsequently two complex quaternion spaces, $\mathbb{H}_g$ and $\mathbb{H}_e$ , can be combined together to become one complex octonion space $\mathbb{O}$. In the complex octonion space $\mathbb{O}$ , the basis vectors are $\textbf{\emph{i}}_j$ and $\textbf{\emph{I}}_j$ , the complex octonion radius vector is, $\mathbb{R} = \mathbb{R}_g + k_{eg} \mathbb{R}_e$ . Herein $k_{eg}$ is one coefficient, to meet the demand of the dimensional homogeneity \cite{weng2}. $r_j$ and $R_j$ are all real. $r_0 = v_0 t$, $v_0$ is the speed of light, and $t$ is the time. $\textbf{\emph{i}}_0 = 1$ . $\textbf{\emph{I}}_j = \textbf{\emph{i}}_j \circ \textbf{\emph{I}}_0$ . $\textbf{\emph{i}}_0 \circ \textbf{\emph{i}}_0 = 1$. $\textbf{\emph{i}}_k \circ \textbf{\emph{i}}_k = - 1$. $\textbf{\emph{I}}_j \circ \textbf{\emph{I}}_j = - 1$. And $i$ is the imaginary unit. $\circ$ denotes the octonion multiplication. $j = 0, 1, 2, 3$. $k = 1, 2, 3$.

Furthermore, in the complex quaternion space $\mathbb{H}_g$ for the gravitational field, the quaternion operator is, $\lozenge = i \textbf{\emph{i}}_0 \partial_0 + \Sigma \textbf{\emph{i}}_k \partial_k$ , with $\partial_j = \partial / \partial r_j$ , and $\nabla = \Sigma \textbf{\emph{i}}_k \partial_k$. By means of the quaternion operator $\lozenge$ and complex octonion quantities, one can explore the physics quantities of electromagnetic and gravitational fields.

\subsection{Field source}

In the complex octonion space $\mathbb{O}$ , the octonion integrating function of field potential is, $\mathbb{X} = \mathbb{X}_g + k_{eg} \mathbb{X}_e$ . $\mathbb{X}_g$ and $\mathbb{X}_e$ are the components of the integrating function of field potential, $\mathbb{X}$ , in the spaces, $\mathbb{H}_g$ and $\mathbb{H}_e$ , respectively. Herein $\mathbb{X}_g = i \textbf{\emph{i}}_0 x_0 + \Sigma \textbf{\emph{i}}_k x_k$. $\mathbb{X}_e = i \textbf{\emph{I}}_0 X_0 + \Sigma \textbf{\emph{I}}_k X_k$. $x_j$ and $X_j$ are all real.

Making use of the integrating function of field potential, $\mathbb{X}$ , the octonion field potential, $\mathbb{A}$ , is able to be defined as (Table 1),
\begin{eqnarray}
\mathbb{A} = i \lozenge^\star \circ \mathbb{X} ~,
\end{eqnarray}
where $\mathbb{A} = \mathbb{A}_g + k_{eg} \mathbb{A}_e$ . $\mathbb{A}_g$ and $\mathbb{A}_e$ are respectively the components of octonion field potential in the subspaces, $\mathbb{H}_g$ and $\mathbb{H}_e$ . The component $\mathbb{A}_g$ is the gravitational potential, while $\mathbb{A}_e$ is the electromagnetic potential. $\mathbb{A}_g = i \textbf{\emph{i}}_0 A_{g0} + \Sigma \textbf{\emph{i}}_k A_{gk}$; $\mathbb{A}_e = i \textbf{\emph{I}}_0 A_{e0} + \Sigma \textbf{\emph{I}}_k A_{ek}$ . $\mathbb{A}_g = i \lozenge^\star \circ \mathbb{X}_g$. $\mathbb{A}_e = i \lozenge^\star \circ \mathbb{X}_e$ . $A_{gj}$ and $A_{ej}$ are all real. $\star$ denotes the complex conjugate.

The octonion field strength, $\mathbb{F}$ , is defined as,
\begin{eqnarray}
\mathbb{F} = \lozenge \circ \mathbb{A}   ~,
\end{eqnarray}
where $\mathbb{F} = \mathbb{F}_g + k_{eg} \mathbb{F}_e$ . $\mathbb{F}_g$ and $\mathbb{F}_e$ are the components of octonion field strength in the subspaces, $\mathbb{H}_g$ and $\mathbb{H}_e$ , respectively. The term $\mathbb{F}_g$ is the gravitational strength, while $\mathbb{F}_e$ is the electromagnetic strength. $\mathbb{F}_g = i \textbf{\emph{i}}_0 F_{g0} + \Sigma \textbf{\emph{i}}_k F_{gk}$; $\mathbb{F}_e = i \textbf{\emph{I}}_0 F_{e0} + \Sigma \textbf{\emph{I}}_k F_{ek}$. $\mathbb{F}_g = \lozenge \circ \mathbb{A}_g$ . $\mathbb{F}_e = \lozenge \circ \mathbb{A}_e$ . $F_{gk}$ and $F_{ek}$ are all complex. $F_{g0}$ and $F_{e0}$ are all real.

From the octonion field strength, it is able to define the octonion field source,
\begin{eqnarray}
\mu \mathbb{S} = - ( i \mathbb{F} / v_0 + \lozenge )^\ast \circ \mathbb{F}   ~,
\end{eqnarray}
where $\mu \mathbb{S} = \mu_g \mathbb{S}_g + k_{eg} \mu_e \mathbb{S}_e - i \mathbb{F}^\ast \circ \mathbb{F} / v_0$ . $\mathbb{S}_g$ and $\mathbb{S}_e$ are respectively the components of field source, $\mathbb{S}$ , in the spaces, $\mathbb{H}_g$ and $\mathbb{H}_e$ . The component $\mathbb{S}_g$ is the gravitational source, while the component $\mathbb{S}_e$ is the electromagnetic source. $\mathbb{S}_g = i \textbf{\emph{i}}_0 S_{g0} + \Sigma \textbf{\emph{i}}_k S_{gk}$; $\mathbb{S}_e = i \textbf{\emph{I}}_0 S_{e0} + \Sigma \textbf{\emph{I}}_k S_{ek}$. $\mu_g \mathbb{S}_g = - \lozenge^\ast \circ \mathbb{F}_g$. $\mu_e \mathbb{S}_e = - \lozenge^\ast \circ \mathbb{F}_e$. The gravitational field coefficient is, $\mu_g < 0$ , and the electromagnetic field coefficient is, $\mu_e > 0$ . $\mu$ , $\mu_g$ , and $\mu_e$ are coefficients. $S_{gj}$ and $S_{ej}$ are all real. $\ast$ denotes the octonion conjugate.

\section{Torque and force}

From the quaternion operator and field source, it is able to define the octonion linear momentum, angular momentum, torque, and force, including the energy, power, mass continuity equation, current continuity equation, force equilibrium equation, and precessional equilibrium equation and so forth (see Ref.[28]).

\subsection{Angular momentum}

From the octonion field source, the octonion linear momentum is defined as,
\begin{eqnarray}
\mathbb{P} = \mu \mathbb{S} / \mu_g   ~,
\end{eqnarray}
where $\mathbb{P} = \mathbb{P}_g + k_{eg} \mathbb{P}_e$ . $\mathbb{P}_g$ and $\mathbb{P}_e$ are respectively the components of octonion linear momentum in the spaces, $\mathbb{H}_g$ and $\mathbb{H}_e$ . $\mathbb{P}_g = i \textbf{\emph{i}}_0 P_{g0} + \Sigma \textbf{\emph{i}}_k P_{gk}$ . $\mathbb{P}_e = i \textbf{\emph{I}}_0 P_{e0} + \Sigma \textbf{\emph{I}}_k P_{ek}$. $\mathbb{P}_g = \mathbb{S}_g - i \mathbb{F}^\ast \circ \mathbb{F} / ( \mu_g v_0 )$ , $\mathbb{P}_e = \mu_e \mathbb{S}_e / \mu_g$ . $P_{gj}$ and $P_{ej}$ are all real.

From the radius vector $\mathbb{R}$ , linear momentum $\mathbb{P}$ , and integrating function of field potential, $\mathbb{X}$ , it is able to define the octonion angular momentum $\mathbb{L}$ as,
\begin{eqnarray}
\mathbb{L} = \mathbb{U}^\star \circ \mathbb{P}   ~,
\end{eqnarray}
where $\mathbb{L} = \mathbb{L}_g + k_{eg} \mathbb{L}_e$ . $\mathbb{L}_g$ and $\mathbb{L}_e$ are the components of octonion angular momentum in two subspaces, $\mathbb{H}_g$ and $\mathbb{H}_e$ , respectively. $\mathbb{L}_g = \mathbb{U}_g^\star \circ \mathbb{P}_g + k_{eg}^2 \mathbb{U}_e^\star \circ \mathbb{P}_e$ . $\mathbb{L}_e = \mathbb{U}_e^\star \circ \mathbb{P}_g + \mathbb{U}_g^\star \circ \mathbb{P}_e$ . $\mathbb{U} = \mathbb{R} + k_{rx} \mathbb{X}$ . $\mathbb{U}_g = \mathbb{R}_g + k_{rx} \mathbb{X}_g$ . $\mathbb{U}_e = \mathbb{R}_e + k_{rx} \mathbb{X}_e$ . $u_j = r_j + k_{rx} x_j$ , $U_j = R_j + k_{rx} X_j$ . $\textbf{u} = \Sigma u_k \textbf{\emph{i}}_k$ . $\textbf{U}_0 = U_0 \textbf{\emph{I}}_0$ . $\textbf{U} = \Sigma U_k \textbf{\emph{I}}_k$. $k_{rx}$ is a coefficient. $x_j$ , $r_j$ , $X_j$ , and $R_j$ are all real.

According to the above, the octonion angular momentum is, $\mathbb{L} = \mathbb{L}_g + k_{eg} \mathbb{L}_e $ . In the complex $S$-quaternion space $\mathbb{H}_e$, the component $\mathbb{L}_e$ includes the conventional electric dipole moment and magnetic dipole moment in the classical field theory. And it can be written as (Table 2),
\begin{eqnarray}
\mathbb{L}_e = \textbf{L}_{e0} + i \textbf{L}_e^i + \textbf{L}_e ~,
\end{eqnarray}
where $\textbf{L}_{e0} = ( u_0 \textbf{P}_{e0} + \textbf{u} \cdot \textbf{P}_e ) + ( p_0 \textbf{U}_0 + \textbf{U} \cdot \textbf{p} )$, $\textbf{L}_e^i = ( - u_0 \textbf{P}_e + \textbf{u} \circ \textbf{P}_{e0} ) + ( - \textbf{U}_0 \circ \textbf{p} + p_0 \textbf{U} )$, $\textbf{L}_e = ( \textbf{u} \times \textbf{P}_e + \textbf{U} \times \textbf{p} )$. The part, $\textbf{L}_{e0}$ , includes the term $\textbf{u} \cdot \textbf{P}_e$ . The part, $\textbf{L}_e^i$ , is considered as the electric dipole moment, covering the term $\textbf{u} \circ \textbf{P}_{e0}$ . The part, $\textbf{L}_e$, is regarded as the magnetic dipole moment, including the term $\textbf{u} \times \textbf{P}_e$ . $p_0 = P_{g0}$. $\textbf{p} = \Sigma P_{gk} \textbf{\emph{i}}_k$. $\textbf{P}_e = \Sigma P_{ek} \textbf{\emph{I}}_k$. $\textbf{P}_{e0} = P_{e0} \textbf{\emph{I}}_0$. $\textbf{L}_{e0} = L_{e0} \textbf{\emph{I}}_0$. $\textbf{L}_e = \Sigma L_{ek} \textbf{\emph{I}}_k$. $\textbf{L}_e^i = \Sigma L^i_{ek} \textbf{\emph{I}}_k$ . $L_{ej}$ and $L^i_{ek}$ are all real.

In the complex quaternion space $\mathbb{H}_g$ , the component $\mathbb{L}_g$ covers the conventional orbital angular momentum in the classical field theory. And it is,
\begin{eqnarray}
\mathbb{L}_g = L_{g0} + i \textbf{L}_g^i + \textbf{L}_g ~ ,
\end{eqnarray}
where $L_{g0} = ( u_0 p_0 + \textbf{u} \cdot \textbf{p} ) + k_{eg}^2 ( \textbf{U}_0 \circ \textbf{P}_{e0} + \textbf{U} \cdot \textbf{P}_e )$ , $\textbf{L}_g = ( \textbf{u} \times \textbf{p} + k_{eg}^2 \textbf{U} \times \textbf{P}_e )$, $\textbf{L}_g^i = ( p_0 \textbf{u} - u_0 \textbf{p} ) + k_{eg}^2 ( \textbf{U} \circ \textbf{P}_{e0} - \textbf{U}_0 \circ \textbf{P}_e )$ . The part, $L_{g0}$ , includes the term, $\textbf{u} \cdot \textbf{p}$ . The part, $\textbf{L}_g^i$ , is similar to the electric dipole moment, covering the term, $p_0 \textbf{u}$ . And the part, $\textbf{L}_g$ , is the orbital angular momentum, which is similar to the magnetic dipole moment, including the term, $\textbf{u} \times \textbf{p}$. $\textbf{L}_g = \Sigma L_{gk} \textbf{\emph{i}}_k$ . $\textbf{L}_g^i = \Sigma L^i_{gk} \textbf{\emph{i}}_k$. $L_{gj}$ and $L^i_{gk}$ are all real.

\subsection{Force}

From the octonion angular momentum, the octonion torque is defined as,
\begin{eqnarray}
\mathbb{W} = - v_0 ( i \mathbb{F} / v_0 + \lozenge ) \circ \mathbb{L}   ~,
\end{eqnarray}
where $\mathbb{W} = \mathbb{W}_g + k_{eg} \mathbb{W}_e$ . $\mathbb{W}_g$ and $\mathbb{W}_e$ are the components of octonion torque in the subspaces, $\mathbb{H}_g$ and $\mathbb{H}_e$ , respectively.
$\mathbb{W}_g = - ( i \mathbb{F}_g + v_0 \lozenge ) \circ \mathbb{L}_g - i k_{eg}^2 \mathbb{F}_e \circ \mathbb{L}_e$. $\mathbb{W}_e = - ( i \mathbb{F}_g + v_0 \lozenge ) \circ \mathbb{L}_e - i \mathbb{F}_e \circ \mathbb{L}_g$. $\mathbb{W}_g = i W_{g0}^i + W_{g0} + i \textbf{W}_g^i + \textbf{W}_g$ . $\mathbb{W}_e = i \textbf{W}_{e0}^i + \textbf{W}_{e0} + i \textbf{W}_e^i + \textbf{W}_e$. $\textbf{W}_g = \Sigma W_{gk} \textbf{\emph{i}}_k$ . $\textbf{W}_g^i = \Sigma W^i_{gk} \textbf{\emph{i}}_k$. $\textbf{W}_{e0}^i = W_{e0}^i \textbf{\emph{I}}_0$. $\textbf{W}_{e0} = W_{e0} \textbf{\emph{I}}_0$. $\textbf{W}_e = \Sigma W_{ek} \textbf{\emph{I}}_k$. $\textbf{W}_e^i = \Sigma W^i_{ek} \textbf{\emph{I}}_k$ . $W_{gj}$ , $W^i_{gj}$, $W_{ej}$ , and $W^i_{ej}$ are all real.

In the space $\mathbb{H}_g$ , the component $\mathbb{W}_g$ can be separated into,
\begin{eqnarray}
\mathbb{W}_g = i W_{g0}^i + W_{g0} + i \textbf{W}_g^i + \textbf{W}_g ~,
\end{eqnarray}
with
\begin{eqnarray}
W_{g0}^i = && ( \textbf{g} \cdot \textbf{L}_g^i / v_0 - \textbf{b} \cdot \textbf{L}_g ) - v_0 ( \partial_0 L_{g0} + \nabla \cdot \textbf{L}_g^i)
\nonumber
\\
&&~~
+ k_{eg}^2 ( \textbf{E} \cdot \textbf{L}_e^i / v_0 - \textbf{B} \cdot \textbf{L}_e ) ~,
\\
W_{g0} = && ( \textbf{b} \cdot \textbf{L}_g^i + \textbf{g} \cdot \textbf{L}_g / v_0 ) - v_0 ( \nabla \cdot \textbf{L}_g )
\nonumber
\\
&&~~
+ k_{eg}^2 ( \textbf{B} \cdot \textbf{L}_e^i + \textbf{E} \cdot \textbf{L}_e / v_0 ) ~,
\\
\textbf{W}_g^i = && ( \textbf{g} \times \textbf{L}_g^i / v_0 - L_{g0} \textbf{b} - \textbf{b} \times \textbf{L}_g )
- v_0 ( \partial_0 \textbf{L}_g + \nabla \times \textbf{L}_g^i )
\nonumber
\\
&&~~
+ k_{eg}^2 ( \textbf{E} \times \textbf{L}_e^i / v_0 - \textbf{B} \circ \textbf{L}_{e0} - \textbf{B} \times \textbf{L}_e ) ~,
\\
\textbf{W}_g = && ( \textbf{g} L_{g0} / v_0 + \textbf{g} \times \textbf{L}_g / v_0 + \textbf{b} \times \textbf{L}_g^i )
\nonumber
\\
&&~~
+ v_0 ( \partial_0 \textbf{L}_g^i - \nabla L_{g0} - \nabla \times \textbf{L}_g )
\nonumber
\\
&&~~
+ k_{eg}^2 ( \textbf{E} \circ \textbf{L}_{e0} / v_0 + \textbf{E} \times \textbf{L}_e / v_0 + \textbf{B} \times \textbf{L}_e^i ) ~,
\end{eqnarray}
where $W_{g0}^i$ is the energy. $-\textbf{W}_g^i$ is the torque. $\textbf{W}_g$ is the curl of angular momentum. The terms, $i \textbf{E} / v_0$ and $\textbf{B}$ , are the components of electromagnetic strength, while two terms, $i \textbf{g} / v_0$ and $\textbf{b}$, are the components of gravitational strength. The electric field intensity is, $\textbf{E} / v_0 = \partial_0 \textbf{A}_e + \nabla \circ \textbf{A}_{e0}$ , while the magnetic flux density is, $\textbf{B} = \nabla \times \textbf{A}_e$ . $\textbf{g} / v_0 = \partial_0 \textbf{A}_g + \nabla A_{g0}$, and $\textbf{b} = \nabla \times \textbf{A}_g$. When $k_p = 2$, the term $W_{g0}^i / 2$ is the energy in the three-dimensional space. Comparing the term $W_{g0}^i / 2$ with the conventional energy in the classical physical theory, one will find that there is, $k_{rx} = 1 / v_0$ . $k_p = (k - 1)$, with $k$ being the dimension of radius vector, $\textbf{r} = \Sigma r_k \textbf{\emph{i}}_k $. $\textbf{A}_{e0} = A_{e0} \textbf{\emph{I}}_0$, $\textbf{A}_e = \Sigma A_{ek} \textbf{\emph{I}}_k$. $\textbf{A}_g = \Sigma A_{gk} \textbf{\emph{i}}_k$. $\textbf{g} = \Sigma g_k \textbf{\emph{i}}_k$ . $\textbf{b} = \Sigma b_k \textbf{\emph{i}}_k$ . $\textbf{E} = \Sigma E_k \textbf{\emph{I}}_k$. $\textbf{B} = \Sigma B_k \textbf{\emph{I}}_k$. $\textbf{W}_g = \Sigma W_{gk} \textbf{\emph{i}}_k$. $\textbf{W}_g^i = \Sigma W^i_{gk} \textbf{\emph{i}}_k$ . $g_k$ , $b_k$, $E_k$ , $B_k$ , $W_{gj}$, and $W^i_{gj}$ are all real.

Similarly, the component $\mathbb{W}_e$ in the space $\mathbb{H}_e$ is separated into,
\begin{eqnarray}
\mathbb{W}_e = i \textbf{W}_{e0}^i + \textbf{W}_{e0} + i \textbf{W}_e^i + \textbf{W}_e ~ ,
\end{eqnarray}
where $\textbf{W}_{e0}$ is the divergence of magnetic dipole moment, $\textbf{W}_e$ is the curl of magnetic dipole moment. $\textbf{W}_{e0}^i = W_{e0}^i \textbf{\emph{I}}_0$. $\textbf{W}_{e0} = W_{e0} \textbf{\emph{I}}_0$ . $\textbf{W}_e = \Sigma W_{ek} \textbf{\emph{I}}_k$ . $\textbf{W}_e^i = \Sigma W^i_{ek} \textbf{\emph{I}}_k$. $W_{ej}$ and $W^i_{ej}$ are all real.

From the octonion torque, the octonion force $\mathbb{N}$ can be defined as (Table 3),
\begin{eqnarray}
\mathbb{N} = - ( i \mathbb{F} / v_0 + \lozenge ) \circ \mathbb{W}  ~,
\end{eqnarray}
where $\mathbb{N} = \mathbb{N}_g + k_{eg} \mathbb{N}_e$ . $\mathbb{N}_g$ and $\mathbb{N}_e$ are the components of force, $\mathbb{N}$ , in the spaces, $\mathbb{H}_g$ and $\mathbb{H}_e$ , respectively. When $k_p = 2$, the term $\textbf{N}_g^i / 2$ is the force in the three-dimensional space. Comparing the term, $\textbf{N}_g^i / 2$ , with the force in the classical field theory states, $k_{eg}^2 = \mu_g / \mu_e < 0$ . $\mathbb{N}_g = - ( i \mathbb{F}_g / v_0 + \lozenge ) \circ \mathbb{W}_g - i k_{eg}^2 ( \mathbb{F}_e / v_0 ) \circ \mathbb{W}_e$. $\mathbb{N}_e = - ( i \mathbb{F}_g / v_0 + \lozenge ) \circ \mathbb{W}_e  - ( i \mathbb{F}_e / v_0 ) \circ \mathbb{W}_g$ . $\mathbb{N}_g = i N_{g0}^i + N_{g0} + i \textbf{N}_g^i + \textbf{N}_g$ . $\mathbb{N}_e = i \textbf{N}_{e0}^i + \textbf{N}_{e0} + i \textbf{N}_e^i + \textbf{N}_e$. $\textbf{N}_g = \Sigma N_{gk} \textbf{\emph{i}}_k$. $\textbf{N}_g^i = \Sigma N^i_{gk} \textbf{\emph{i}}_k$. $\textbf{N}_{e0}^i = N_{e0}^i \textbf{\emph{I}}_0$ . $\textbf{N}_{e0} = N_{e0} \textbf{\emph{I}}_0$ . $\textbf{N}_e = \Sigma N_{ek} \textbf{\emph{I}}_k$. $\textbf{N}_e^i = \Sigma N^i_{ek} \textbf{\emph{I}}_k$ . $N_{gj}$, $N^i_{gj}$ , $N_{ej}$ , and $N^i_{ej}$ are all real.

The component $\mathbb{N}_g$ situates in the complex quaternion space $\mathbb{H}_g$ , and the component $\mathbb{N}_e$ locates in the complex $S$-quaternion space $\mathbb{H}_e$ . Further, the component $\mathbb{N}_g$ can be expressed as,
\begin{equation}
\mathbb{N}_g = \emph{i} N_{g0}^i + N_{g0} + \emph{i} \textbf{N}_g^i + \textbf{N}_g ~,
\end{equation}
with
\begin{eqnarray}
N_{g0}^i = && ( \textbf{g} \cdot \textbf{W}_g^i / v_0 - \textbf{b} \cdot \textbf{W}_g ) / v_0 - ( \partial_0 W_{g0} + \nabla \textbf{W}_g^i )
\nonumber
\\
&&~~
+ k_{eg}^2 ( \textbf{E} \cdot \textbf{W}_e^i / v_0 - \textbf{B} \cdot \textbf{W}_e ) / v_0 ~,
\\
N_{g0} = && ( \textbf{g} \cdot \textbf{W}_g / v_0 + \textbf{b} \cdot \textbf{W}_g^i ) / v_0 + ( \partial_0 W_{g0}^i - \nabla \cdot \textbf{W}_g )
\nonumber
\\
&&~~
+ k_{eg}^2 ( \textbf{E} \cdot \textbf{W}_e / v_0 + \textbf{B} \cdot \textbf{W}_e^i ) / v_0 ~,
\\
\textbf{N}_g^i = && ( W_{g0}^i \textbf{g} / v_0 + \textbf{g} \times \textbf{W}_g^i / v_0 - W_{g0} \textbf{b} - \textbf{b} \times \textbf{W}_g ) / v_0
\nonumber
\\
&&~~
+ k_{eg}^2 ( \textbf{E} \circ \textbf{W}_{e0}^i / v_0 + \textbf{E} \times \textbf{W}_e^i / v_0 - \textbf{B} \circ \textbf{W}_{e0} - \textbf{B} \times \textbf{W}_e ) / v_0
\nonumber
\\
&&~~
- ( \partial_0 \textbf{W}_g + \nabla W_{g0}^i + \nabla \times \textbf{W}_g^i )
~,
\\
\textbf{N}_g = && ( W_{g0} \textbf{g} / v_0 + \textbf{g} \times \textbf{W}_g / v_0 + W_{g0}^i \textbf{b} + \textbf{b} \times \textbf{W}_g^i ) / v_0
\nonumber
\\
&&~~
+ k_{eg}^2 ( \textbf{E} \circ \textbf{W}_{e0} / v_0 + \textbf{E} \times \textbf{W}_e/ v_0 + \textbf{B} \circ \textbf{W}_{e0}^i + \textbf{B} \times \textbf{W}_e^i ) / v_0 \nonumber
\\
&&~~
+ ( \partial_0 \textbf{W}_g^i - \nabla W_{g0} - \nabla \times \textbf{W}_g )
~,
\end{eqnarray}
where $N_{g0}^i$ is the torque divergence. $N_{g0}$ is the power density. $\textbf{N}_g^i$ is the force density. $\textbf{N}_g$ is the torque derivative. The energy gradient, $(- \nabla W_{g0}^i)$ , is able to exert an influence on various particles, including the neutral and charged particles.

\begin{table}[h]
\caption{Comparison between the orbital angular momentum with the electromagnetic moment, in the octonion angular momentum.}
\center
\begin{tabular}{@{}ll@{}}
\hline\hline
Angular~momentum~term                                       &  Electromagnetic~moment                                        \\
\hline
$L_{g0}$ , term                                             &  $\textbf{L}_{e0}$ , term                                      \\
$\textbf{L}_g^i$ , term                                     &  $\textbf{L}_e^i$ , electric dipole moment                     \\
$\textbf{L}_g$ , orbital angular momentum                   &  $\textbf{L}_e$ , magnetic dipole moment                       \\
\hline\hline
\end{tabular}
\end{table}

\section{Angular velocity of precession}

According to the quaternion torque, $\mathbb{W}_g$ , there are the interacting torque and energy between the magnetic dipole moment, $\textbf{L}_e$ , with the external magnetic field. But there is only the interacting torque between the term, $\textbf{L}_{e0}$ , with the external magnetic field. Each of these terms interacting with the external magnetic field is capable of inducing the angular velocity of precession, generating a few new components of magnetic dipole moment, altering the interacting energy between each component of magnetic dipole moment with the external magnetic field, resulting in the new energy gradient finally.

When $\mathbb{N}_g = 0$ , there are the force equilibrium equation, $\textbf{N}_g^i = 0$ , and the precessional equilibrium equation, $\textbf{N}_g = 0$ . From the latter, it is able to deduce the angular velocity of precession, for the charged particle in the external magnetic field. In case there is only the magnetic flux density $\textbf{B}$ , the precessional equilibrium equation, $ \textbf{N}_g = 0 $, can be reduced into,
\begin{eqnarray}
k_{eg}^2 ( \textbf{B} \circ \textbf{W}_{e0}^i + \textbf{B} \times \textbf{W}_e^i ) / v_0 + ( \partial_0 \textbf{W}_g^i - \nabla W_{g0} - \nabla \times \textbf{W}_g ) = 0 ~,
\end{eqnarray}
where $\textbf{W}_{e0}^i \approx k_p \textbf{P}_{e0} v_0$ , $\textbf{W}_g^i \approx k_{eg}^2 ( - \textbf{B} \circ \textbf{L}_{e0} - \textbf{B} \times \textbf{L}_e) $, and $\textbf{W}_g \approx k_p \textbf{p} v_0$ .

1) When there are only two terms, $ ( k_{eg}^2 \textbf{B} \circ \textbf{W}_{e0}^i / v_0 ) $ and $\nabla \times \textbf{W}_g$ , the above will be reduced to
\begin{eqnarray}
 - v_0 \nabla \times \textbf{p} + k_{eg}^2 \textbf{B} \circ \textbf{P}_{e0}  = 0  ~,
\end{eqnarray}
consequently, for the charged particle in the external magnetic fields, the angular frequency of Larmor precession is,
\begin{eqnarray}
\omega = - B q V_0 / ( k m v_0 ) ~,
\end{eqnarray}
where $k_{eg}^2 \textbf{P}_{e0} = q V_0 \emph{\textbf{I}}_0$ , and $\textbf{p} = m \textbf{v}$ , for a charged particle. The velocity, $\textbf{v} = \Sigma v_k \textbf{\emph{i}}_k$, in the space $\mathbb{H}_g$ , can be rewritten as, $\textbf{v} = \overrightarrow{\omega} \times \textbf{r}$ , with $\overrightarrow{\omega}$ being the angular velocity. $\omega$ and $B$ are the magnitude of the angular velocity $\overrightarrow{\omega}$ and $\textbf{B}$ respectively, with $ \nabla \times \textbf{v} = k \overrightarrow{\omega} $ . $m$ is the density of inertial mass. $v_k$ is real. $V_0 / v_0 \approx 1$ .

It reveals that there is still the angular velocity of precession for a charged particle, even if under the extreme condition that there is not any existing torque term, that is, $\textbf{W}_g^i = 0$ .

2) When there are merely two terms, $ \partial_0 ( k_{eg}^2 \textbf{B} \circ \textbf{L}_{e0} ) $ and $\nabla \times \textbf{W}_g$ , Eq.(21) is degenerated to
\begin{eqnarray}
 - v_0 k_p \nabla \times \textbf{p} + \partial_0 ( - k_{eg}^2 \textbf{B} \circ \textbf{L}_{e0} )  = 0  ~,
\end{eqnarray}
subsequently the angular frequency of precession for the charged particle is,
\begin{eqnarray}
\omega =  B q V_0 / ( k k_p m v_0 ) ~ ,
\end{eqnarray}
where $\textbf{L}_{e0} \approx r_0 \textbf{P}_{e0} $ .

It is easy to find that the angular frequency of precession may become much more complicated than that in Eqs.(23) and (25), when there are two terms, $ ( k_{eg}^2 \textbf{B} \circ \textbf{W}_{e0}^i / v_0 ) $ and $ \partial_0 ( - k_{eg}^2 \textbf{B} \circ \textbf{L}_{e0} ) $ , simultaneously.

3) In case there are only two terms, $ \partial_0 ( k_{eg}^2 \textbf{B} \times \textbf{L}_e ) $ and $\nabla \times \textbf{W}_g$ , Eq.(21) is simplified to
\begin{eqnarray}
 - v_0 k_p \nabla \times \textbf{p} + \partial_0 ( - k_{eg}^2 \textbf{B} \times \textbf{L}_e ) = 0  ~,
\end{eqnarray}
further, it will yield the angular frequency of precession, $ \omega $ , which is connected to the time-dependent magnetic field or alterable magnetic dipole moment.

Moreover, the angular frequency of precession may be impacted by some other arguments, including the gravitational strength, electromagnetic strength, and a few terms of octonion torque $\mathbb{W}$ . So the angular frequency of precession may be complicated enough, when there exist several arguments simultaneously.

\begin{table}[h]
\caption{Some physical quantities relevant to the electromagnetic and gravitational fields, in the complex octonion space.}
\center
\begin{tabular}{@{}ll@{}}
\hline\hline
Octonion~physics~quantity ~~~~~~~~    &   Definition                                                                                  \\
\hline
field~potential                       &  $\mathbb{A} = i \lozenge^\star \circ \mathbb{X}  $                                           \\
field~strength                        &  $\mathbb{F} = \lozenge \circ \mathbb{A}  $                                                   \\
field~source                          &  $\mu \mathbb{S} = - ( i \mathbb{F} / v_0 + \lozenge )^* \circ \mathbb{F} $                   \\
linear~momentum                       &  $\mathbb{P} = \mu \mathbb{S} / \mu_g $                                                       \\
angular~momentum                      &  $\mathbb{L} = \mathbb{U}^\star \circ \mathbb{P} $                                            \\
octonion~torque                       &  $\mathbb{W} = - v_0 ( i \mathbb{F} / v_0 + \lozenge ) \circ \mathbb{L} $                     \\
octonion~force                        &  $\mathbb{N} = - ( i \mathbb{F} / v_0 + \lozenge ) \circ \mathbb{W} $                         \\
\hline\hline
\end{tabular}
\end{table}

\section{Angular momentum operator}

In the complex octonion space $\mathbb{O}$ , by means of the properties of octonions, it is able to represent the octonion as the exponential form, defining the wave function of octonion physical quantity. Making use of the wave function relevant to the octonion angular momentum, one can define the angular momentum operator, deducing the quantization results regarding major constituents of the orbital angular momentum. Further it is capable of defining the octonion wave equation, which can be reduced to the Dirac wave equation and Schr\"{o}dinger wave equation.

\subsection{Wave function}

In the complex quaternion space $\mathbb{H}_g$ , the quaternion quantity, $\mathbb{Y}_g = y_0 + \Sigma y_k \textbf{\emph{i}}_k$ , possesses the scalar part, $y_0$ , and vector part, $\Sigma y_k \textbf{\emph{i}}_k$. According to the quaternion property, the quaternion $\mathbb{Y}_g$ can be written as, $\mathbb{Y}_g = y_g exp( \alpha_g \textbf{\emph{i}}_g )$ . Further, in case $\mid y_0 / y_g \mid \ll 1$, the angle, $\alpha_g = arc cos ( y_0 / y_g )$, can be expanded in a Taylor series, $\alpha_g \approx \pi / 2 - ( y_0 / y_g )$ . As a result, one can define a wave function, $\Psi_g = - \textbf{\emph{i}}_g \circ \mathbb{Y}_g$ , relevant to the quaternion $\mathbb{Y}_g$ , that is,
\begin{eqnarray}
\Psi_g = y_g exp \{ - ( y_0 / y_g ) \textbf{\emph{i}}_g \}  ~,
\end{eqnarray}
where $\textbf{\emph{i}}_g = \Sigma y_k \textbf{\emph{i}}_k / ( \Sigma y_k^2 )^{1/2}$ . $y_g = ( \Sigma y_j^2 )^{1/2}$ . $cos \alpha_g = y_0 / y_g$ . $\textbf{\emph{i}}_g$ is an unit vector, with $\textbf{\emph{i}}_g^2 = - 1$. $y_j$ is real.

The octonion can be considered as an ordered couple of two quaternions. Therefore the research methods for the exponential form and wave function can be extended from the quaternion space into the octonion space. In other words, the octonion can be expressed as the exponential form also, defining the wave function regarding the octonion physical quantity (Appendices A and B).

In the complex octonion space $\mathbb{O}$ , it is able to define the wave function, $\Psi_L = \mathbb{L} / \hbar$ , from the octonion angular momentum, $\mathbb{L} = \mathbb{L}_g + k_{eg} \mathbb{L}_e$ . This dimensionless wave function, $\Psi_L$, relates with not only the quaternion angular momentum $\mathbb{L}_g$ , but also the $S$-quaternion electromagnetic moment $\mathbb{L}_e$. In the following context, we discuss merely a simple case, that is, considering only the contribution of the angular momentum $\mathbb{L}_g$ to the wave function $\Psi_L$ . In other words, when the contribution of electromagnetic moment $\mathbb{L}_e$ can be neglected, the octonion wave function $\Psi_L$ will be degenerated to the quaternion wave function, $\Psi_{Lg} = \mathbb{L}_g / \hbar$.

Making use of the quaternion property, the quaternion wave function $\Psi_{Lg}$ is expressed as, $\Psi_{Lg} = i L_9 exp \{ \textbf{\emph{i}}_9 ( \pi / 2 ) \} + L_q exp ( \textbf{\emph{i}}_q \alpha_q )$ , in the complex quaternion space $\mathbb{H}_g$ . In case there are, $\mathbb{X}_g = 0$, $ \Sigma ( L_{gj}^i )^2 \ll \Sigma ( L_{gj} )^2$ , and $\mid L_{g0} \mid / \{ \Sigma ( L_{gj} )^2 \}^{1/2} \ll 1$, the angle $\alpha_q$ will be reduced to the angle $\alpha_q^\prime$ . Meanwhile the wave function $\Psi_{Lg}$ is degenerated approximately to the wave function,
\begin{eqnarray}
\Psi_{Lg}^\prime = L_q \textbf{\emph{i}}_q \circ exp ( - \textbf{\emph{i}}_q \alpha_q^\prime )  ~,
\end{eqnarray}
where $\textbf{L}_g^i / \hbar = L_9 exp \{ \textbf{\emph{i}}_9 ( \pi / 2 ) \}$ , $( L_{g0} + \textbf{L}_g ) / \hbar = L_q exp ( \textbf{\emph{i}}_q \alpha_q )$ . $\alpha_q$ and $\alpha_q^\prime$ both are real. $\textbf{\emph{i}}_9$ and $\textbf{\emph{i}}_q$ both are unit vectors. $\textbf{\emph{i}}_9^2 = - 1$. $\textbf{\emph{i}}_q^2 = - 1$.

By means of the coordinate transformation, $ \Psi = - \textbf{\emph{i}}_q \circ \Psi_{Lg}^\prime$ , one can obtain the wave function $\Psi$ regarding the angular momentum $\mathbb{L}_g$ , and
\begin{eqnarray}
\Psi = L_q exp ( - \textbf{\emph{i}}_q \alpha_q^\prime )  ~,
\end{eqnarray}
further, from the coordinate transformation, $\Psi^{\prime\prime} = - \textbf{z} \circ \Psi_{Lg}^\prime$ , with $L_q = 1$, it is able to achieve another wave function $\Psi^{\prime\prime}$ regarding the angular momentum $\mathbb{L}_g$ ,
\begin{eqnarray}
\Psi^{\prime\prime} = (- \textbf{z} ) \circ \{ \textbf{\emph{i}}_q \circ exp ( - \textbf{\emph{i}}_q \alpha_q^\prime ) \} ~,
\end{eqnarray}
where $\textbf{z} = \Sigma z_k \textbf{\emph{i}}_k$ , $z_k$ is a dimensionless real number. Especially, when the direction of unit vector $\textbf{\emph{i}}_q$ is incapable of playing a major role in the wave function, the unit vector $\textbf{\emph{i}}_q$ can be replaced by the imaginary unit, $i$ . So the wave function $\Psi^{\prime\prime}$ can be degenerated to one wave function in the Quantum Mechanics.

The above reveals what plays an important role in the quantum theory is the wave function, $\Psi^{\prime\prime} = - \textbf{z} \circ \Psi_{Lg}^\prime$ , rather than $\Psi_{Lg}^\prime$ . Generally, from the wave function $\Psi_L$ , it is able to define an octonion wave function,
\begin{eqnarray}
\Psi_{ZL} = - \mathbb{Z}_L \circ \Psi_L ~,
\end{eqnarray}
where $\mathbb{Z}_L$ is an auxiliary quantity, and also an octonion dimensionless quantity, including $\textbf{\emph{i}}_g$ and $\textbf{z}$ . $\Psi_{ZL}$ covers $\Psi$ and $\Psi^{\prime\prime}$ and so on.

\subsection{Orbital angular momentum}

In the complex quaternion space $\mathbb{H}_g$ , in terms of the physical properties of wave functions, $\Psi$ and $\Psi^{\prime\prime}$ , the linear momentum operator $\hat{p}_j$ is defined as,
\begin{eqnarray}
\hat{p}_k = - \textbf{\emph{i}}_q \hbar \partial / \partial u_k ~ , ~~  \hat{p}_0 = \textbf{\emph{i}}_q \hbar \partial / \partial u_0    ~,
\end{eqnarray}
where the linear momentum operator $\hat{p}_j$ corresponds to the component $P_{gj}$ of linear momentum $\mathbb{P}_g$ . Making use of the eigenfunction property, the linear momentum operator is able to meet the demand of the photon energy, $E^\prime = h \nu^\prime$, and the wavelength of matter wave, $\lambda^\prime = h / p^\prime$ , in the Quantum Mechanics. $E^\prime$ is the energy, $\nu^\prime$ is the frequency. $\lambda^\prime$ is the wavelength, $p^\prime$ is the linear momentum.

Further, the angular momentum operator $\hat{L}_{gk}$ is defined as,
\begin{eqnarray}
&& \hat{L}_{g1} = - \textbf{\emph{i}}_q \hbar ( u_2 \partial / \partial u_3 - u_3 \partial / \partial u_2 )  ~,
\\
&& \hat{L}_{g2} = - \textbf{\emph{i}}_q \hbar ( u_3 \partial / \partial u_1 - u_1 \partial / \partial u_3 )  ~,
\\
&& \hat{L}_{g3} = - \textbf{\emph{i}}_q \hbar ( u_1 \partial / \partial u_2 - u_2 \partial / \partial u_1 )  ~,
\end{eqnarray}
where the operator $\hat{L}_{gk}$ corresponds to the component $L_{gk}$ of major constituent, $\textbf{u} \times \textbf{p}$ , of the orbital angular momentum $\textbf{L}_g$ .

When the direction property of unit vector $\textbf{\emph{i}}_q$ can be neglected approximately, the unit vector $\textbf{\emph{i}}_q$ will be replaced by the imaginary unit, $i$ . In this circumstance, substituting the coordinate value $u_j$ for the coordinate value $r_j$, the quantization approaches for the orbital angular momentum, in the Quantum Mechanics, can be extended into the above equations, achieving the quantization results about the orbital angular momentum described with the quaternions.

The above means that the quantization results of the major constituent, $\textbf{u} \times \textbf{p}$, of the orbital angular momentum $\textbf{L}_g$ , under certain circumstances in the complex quaternion space $\mathbb{H}_g$ , are similar to that in the Quantum Mechanics.

\subsection{Wave equation}

In the complex octonion space $\mathbb{O}$ , from the wave function $\Psi_{ZL}$ , we can define a wave function, $\Psi_{WL}$ , regarding the octonion torque,
\begin{eqnarray}
\Psi_{WL} = - \{ i \mathbb{W}^\star + \hbar v_0 \lozenge \} \circ \Psi_{ZL}   ~,
\end{eqnarray}
further, when $\Psi_{WL} = 0$, the Dirac wave equation in the complex octonion space can be deduced from the above (Appendices C and D). Under certain circumstances, the Dirac wave equation described with the octonions will be degenerated to,
\begin{eqnarray}
&& \{ i \hbar v_0 \partial_0 - k_{eg}^2 \textbf{A}_{e0} \circ \textbf{P}_{e0} + \hbar v_0 \nabla
\nonumber
\\
&&~~~~~~~~~~~~~~
+ i k_{eg}^2 \textbf{A}_e \circ \textbf{P}_{e0} + k_p m^\prime v_0^2 \} \circ \Psi_{ZL} = 0   ~,
\end{eqnarray}
where $m^\prime$ is the density of gravitational mass.

In the complex octonion space $\mathbb{O}$ , acting an appropriate operator $\lozenge_{DS}$ on the wave function $\Psi_{WL}$ , it is able to achieve a correlative wave function $\Psi_{WL}^\prime$ ,
\begin{eqnarray}
\Psi_{WL}^\prime = \lozenge_{DS} \circ \Psi_{WL}    ~,
\end{eqnarray}
where $\lozenge_{DS} = i \mathbb{W}_{DS} / ( \hbar v_0 ) + \lozenge$ , and $\mathbb{W}_{DS} = \mathbb{W}^\star + i ( 2 k_p m^\prime v_0^2 )$ .

When $\Psi_{WL}^\prime = 0$, the Schr\"{o}dinger wave equation in the complex octonion space can be derived from the above (Appendix E). Under certain circumstances, the wave equation described with the octonions will be degenerated to,
\begin{eqnarray}
E^\prime \Psi (\textbf{r}) = && \{[ \textbf{W}_g - i ( \hbar v_0 ) \nabla ] \cdot [ \textbf{W}_g - i ( \hbar v_0 ) \nabla ] / ( 2 k_p m^\prime v_0^2 ) - E_w
\nonumber
\\
&&~~~~
+ i k \hbar \textbf{b} / ( 2 k_p ) + i k q V_0 \hbar \textbf{\emph{I}}_0 \circ \textbf{B} / ( 2 k_p m^\prime v_0 ) \} \circ \Psi (\textbf{r})    ~,
\end{eqnarray}
where the wave function, $\Psi^{\prime\prime} = \Psi (\textbf{r}) exp ( - i E_s t / \hbar )$ , is a major constituent of wave function $\Psi_{ZL}$ , in case the unit vector $\textbf{\emph{i}}_q$ can be replaced by the imaginary unit $i$. $E_s = k_p m^\prime v_0^2 + E^\prime$. $W_{g0}^i = k_p m^\prime v_0^2 + E_w$ . When last two terms can be neglected, the above will be degenerated to the Schr\"{o}dinger wave equation in the Quantum Mechanics. $q$ is the density of electric charge.

In the dimensional analysis, the dimension of the above equation is that of the energy or torque. According to the octonion torque $\mathbb{W}$ , the last two terms in the above are the vector terms. That is, the last two terms in the above are the torque terms or other vector terms, rather than the energy terms, in the complex octonion space. Therefore the last term in the above cannot be applied to explain the spin phenomena, nor reveal the origination of spin angular momentum.

Presenting a striking contrast to the above equation is that the precessional equilibrium equation, $\textbf{N}_g = 0$ , is able to account for the origination of spin angular momentum, in the complex octonion space.

\section{Spin angular momentum}

In the complex octonion space, the force equilibrium equation, $\textbf{N}_g^i = 0$ , can deduce the angular velocity of revolution for a charged particle. Meanwhile the precessional equilibrium equation, $\textbf{N}_g = 0$, is able to infer the angular velocity of precession for a charged particle. Apparently, each component of angular momentums, caused by either the revolution or precession, would be a part of the orbital angular momentum $\textbf{L}_g$ . And each component of magnetic moments, caused by either the revolution or precession, would be a part of the magnetic dipole moment $\textbf{L}_e$ . The study reveals that a portion of precessional motions can be considered as the spin motions, explaining the origination, quantum number, orientation, and attenuation of the spin angular momentum (Table 4).

\subsection{Origination}

In the complex octonion space, according to the precessional equilibrium equation, $\textbf{N}_g = 0$, the existence of angular momentum is not a prerequisite for generating the precession. From Eq.(21), each of the electric field, magnetic field, gravitational field, and torque derivative may become an immediate cause to induce the precession, for a charged particle. When there is no angular momentum in the electromagnetic field, some terms (such as, $\textbf{B} \circ \textbf{W}_{e0}^i$) can still induce the precession motions. This is called as the first-type precession temporarily.

The above means that a few arguments are capable of inducing the first-type precession in the electromagnetic fields, even if there is no angular momentum. The first-type precession may produce the components of orbital angular momentum and magnetic dipole moment, yielding the ingredients of interacting energy and torque, in the electromagnetic and gravitational fields. According to Eq.(23), a portion of precessional motions can be regarded as the spin motions in the Quantum Mechanics. Obviously, the existence of electromagnetic field exerts an influence on the spin motion.

When there are the components of angular momentum, some influence factors (such as, torque derivative) will induce the precession, according to the precessional equilibrium equation, $\textbf{N}_g = 0$ . It is called as the second-type precession contemporarily. In the electromagnetic and gravitational fields, the second-type precession will also generate a few new components of orbital angular momentum, magnetic dipole moment, interacting energy, and interacting torque.

Apparently, the first-type precession covers the spin motion in the Quantum Mechanics. Meanwhile the second-type precession is different to the conventional gyroscopic precession in the Theoretical Mechanics. Because the latter meets the requirement of Eq.(8). In contrast to the second-type precession, the first-type precession is much more difficult to be observed and measured.

\subsection{Angular velocity}

In a stationary and uniform magnetic field $\textbf{B}$ , the force equilibrium equation, $\textbf{N}_g^i = 0$ , is able to infer the angular speed of revolution, $\omega_r = - B q V_0 / (m v_0)$ , for a charged particle. Meanwhile the precessional equilibrium equation, $\textbf{N}_g = 0$ , is capable of deducing the angular speed of precession, $\omega_p = - B q V_0 / (k m v_0)$. In the planar particle motion, there is, $\omega_r / \omega_p = 2$ , for $k = 2$. In other words, the external magnetic field $\textbf{B}$ induces not only the angular velocity of revolution, but also the angular velocity of precession. When $k = 2$, the induced component of angular momentum, related with the precession, can be considered as the component of spin angular momentum in the Quantum Mechanics.

In the quantum theory described with the complex octonion, the component of angular momentum, caused by the precession angular speed $\omega_p$ , is essentially the same as that caused by the revolution angular speed $\omega_r$ . Both of them belong to the orbital angular momentum $\textbf{L}_g$ . So their quantization rules are identical, for these two components of angular momentums. In the magnetic field $\textbf{B}$ , the component of angular momentum, $\textbf{L}_{g(r)}$ , relevant to the angular speed $\omega_r$ , is able to be quantized, and the quantum number is $l$ . The component of angular momentum, $\textbf{L}_{g(p)}$ , relevant to the angular speed $\omega_p$ , can be quantized also, and the quantum number is $l /2$ , for $k = 2$ (Appendix F).

In certain comparatively complicated electromagnetic fields, the precessional equilibrium equation, $\textbf{N}_g = 0$ , will deduce some intricate angular speeds of precession. These precession angular speeds may be diversiform, and even they could be mistaken for the background noises, which are difficult to be eliminated. The quantum results for these components of angular momentum will be involuted, occupying various quantum numbers, especially the half quantum number.

\begin{table}[h]
\caption{A part of main physical properties relevant to the spin angular momentum, in the complex octonion space. Some precessional motions can be regarded as the spin motions.}
\center
\begin{tabular}{@{}ll@{}}
\hline\hline
Terms                                                       &  Physical~properties                                                                       \\
\hline
Precessional motion                                         &  The angular velocity of precession can be derived                                         \\
                                                            &  ~~~~from the precessional equilibrium equation, $\textbf{N}_g = 0$ .                      \\
Origination                                                 &  Each of electric field, magnetic field, gravitational                                     \\
                                                            &  ~~~~field, and torque derivative and so on is still                                       \\
                                                            &  ~~~~able to induce the precession, even if there is                                       \\
                                                            &  ~~~~no angular momentum, according to $\textbf{N}_g = 0$.                                 \\
Ratio of angular speeds                                     &  A ratio of the angular speed of precession to the                                         \\
                                                            &  ~~~~angular speed of revolution is multiple-value,                                        \\
                                                            &  ~~~~including 1, 1/2, and 1/3 and so forth.                                               \\
Orientation                                                 &  When there is only the term, $\textbf{B} \circ \textbf{W}_{e0}^i$ , in the                \\
                                                            &  ~~~~magnetic field, the induced angular speed of                                          \\
                                                            &  ~~~~precession is $\omega_p$ , meanwhile the direction of                                 \\
                                                            &  ~~~~precessional axis orients the $\textbf{B} \circ\textbf{\emph{I}}_0$ .                 \\
Attenuation                                                 &  If the external magnetic field is taken away, the                                         \\
                                                            &  ~~~~spin motion of a charged particle is gradually                                        \\
                                                            &  ~~~~faded away. The generation, attenuation,                                              \\
                                                            &  ~~~~and reverse of spin angular momenta, caused                                           \\
                                                            &  ~~~~by external factors, may take a little time,                                          \\
                                                            &  ~~~~including the relaxation and spin-flip times.                                         \\
\hline\hline
\end{tabular}
\end{table}

\subsection{Orientation}

The precessional equilibrium equation, $\textbf{N}_g = 0$ , is able to determine not only the angular speed of precession, but also the orientation of precession. In the electromagnetic field, the direction of the angular velocity of precession, induced by the electromagnetic strength, is related to the direction of electromagnetic strength. Therefore, the angular velocity of precession may orient a certain direction for a period of time. The direction will remain orienting, until the disappearance of electromagnetic strength or the intervention of other influence factors, resulting in the modification of precession direction, and even the reversal of precession direction or the spin flip.

In the magnetic field, the magnetic flux density $\textbf{B}$ can induce a new precession angular velocity and a new component, $\textbf{L}_{g(p)(new)}$ , of orbital angular momentum, even if the existing orbital angular momentum, $\textbf{L}_{g(existing)}$ , is zero. When there is only the term, $\textbf{B} \circ \textbf{W}_{e0}^i$ , the induced angular speed of precession is, $\omega_p = - B q V_0 / (k m v_0)$ , and the orientation of precession axis is $\textbf{B} \circ\textbf{\emph{I}}_0$. Further, in virtue of the contributions of other interacting torque terms, the orientations of precession axis of charged particles, in an ion beam, will either turn towards or depart from the direction, $\textbf{B} \circ\textbf{\emph{I}}_0$ , finally. When the direction of external magnetic field $\textbf{B}$ is altered, the orientation of precession axis will be changed accordingly.

Similarly, in the stationary electric field, the precessional equilibrium equation, $\textbf{N}_g = 0$, can also infer the angular speed of precession and the orientation of precession, for the charged particles. Further, in the time-dependent electric/magnetic field, it is able to induce more complicated angular speeds of precession and the orientations of precession. Because the time-dependent electromagnetic field will encounter more influence factors.

\subsection{Attenuation}

According to the precessional equilibrium equation, $\textbf{N}_g = 0$ , the external electromagnetic field and torque derivative and so forth are capable of deducing new angular velocities of precession, producing the new component, $\textbf{L}_{g(p)(new)}$ , of angular momentum and the new component, $\textbf{L}_{e(p)(new)}$ , of magnetic dipole moment. In other words, two new components, $\textbf{L}_{g(p)(new)}$ and $\textbf{L}_{e(p)(new)}$ , are originated from the existence of the external electromagnetic field and torque derivative and so on. After neglecting some tiny components, there is a certain proportion relation between the major constituent of $\textbf{L}_{g(p)(new)}$ with that of $\textbf{L}_{e(p)(new)}$.

In the precessional equilibrium equation, $\textbf{N}_g = 0$ , various torque derivatives, electromagnetic fields, and gravitational fields will induce several complicated angular velocities of precession, increasing multiform components of angular momentum and magnetic dipole moment. These recent increasing components of angular momentum will still satisfy the existing quantization rules for the angular momentum. As a result, these recent increasing components of angular momentum and magnetic dipole moment will be seized of diversiform quantum numbers. In other words, besides two dominating quantum numbers, $l$ and $l / 2$ , there are other kinds of quantum numbers.

With the passage of time, the recent increasing spin angular momentum will gradually melt into the existing orbital angular momentum, while the spin angular momentum is declining. Moreover the existence of external electromagnetic field and torque derivative and so forth will alter the spin angular momentum, and even decline or reverse it. It means that these vulnerable angular velocities of precession will be easily interfered by some other external factors. The generation, attenuation, and reverse of the spin angular momentum, caused by external factors, may take a little time, including the relaxation and spin-flip times.

The generating of spin angular momentum is similar to that of counterforce in the classical mechanics. When the external force exerts an influence on an object, it is able to induce a counterforce accordingly. If the external force is removed, the counterforce will be vanished quickly. In a similar way to the counterforce, in case the external magnetic field exerts an influence on a charged particle, it is capable of inducing the spin motion of a charged particle accordingly. On condition that the external magnetic field is taken away, the spin motion of a charged particle will be gradually faded away. So the spin motion is a sort of suppressed precession.

The above states that the so-called spin motion in the Quantum Mechanics is merely a precessional motion in the complex octonion space, according to the physical properties of precession, quantum number, and orientation of spin angular momentum. Furthermore, the precessional motion induced by the term, $\textbf{B} \circ \textbf{W}_{e0}^i$ , is the major constituent of spin motion in the Quantum Mechanics. It can be applied to explain some physical phenomena in the Stern-Gerlach and Zeeman experiments and so forth.

Obviously the above inferences will be beneficial to give an insight into the total angular momentum in the proton spin puzzle.

\begin{table}[h]
\caption{The octonion angular momentum of proton can be separated into six components, in the complex octonion space. According to the Section 6, in the complex octonion space, the spin angular momentum belongs to the orbital angular momentum, obeying the same quantization rules with different quantum numbers, meanwhile the spin magnetic moment is in the possession of the magnetic dipole moment.}
\center
\begin{tabular}{@{}ll@{}}
\hline\hline
Component                                             &  Physical meaning                                                \\
\hline
$L_{g0}$                                              &  scalar term                                                     \\
$\textbf{L}_g^i$                                      &  vector term                                                     \\
$\textbf{L}_g$                                        &  orbital angular momentum,                                       \\
                                                      &  ~~~~including the spin angular momentum                         \\
$\textbf{L}_{e0}$                                     &  scalar-like term                                                \\
$\textbf{L}_e^i$                                      &  electric dipole moment                                          \\
$\textbf{L}_e$                                        &  magnetic dipole moment,                                         \\
                                                      &  ~~~~including the spin magnetic moment                          \\
\hline\hline
\end{tabular}
\end{table}

\section{Proton spin puzzle}

In the complex octonion space, the precession angular speed and orientation can be deduced from the precessional equilibrium equation, $\textbf{N}_g = 0$ , for the proton in the electromagnetic fields. In the two-dimensional stationary magnetic field, the spin angular momentum of the proton may possess a half quantum number, and orient over a period of time. However, as a part of precession, the spin motion of the proton may be very complicated and volatile, in an arbitrary external electromagnetic field. Especially, the induced spin angular momentum must even be quite intricate, within the proton (Table 5).

\subsection{Proton spin}

According to the precessional equilibrium equation, $\textbf{N}_g = 0$ , when the external electromagnetic fields and torque derivative exert an influence on the proton, it may induce some new terms of spin angular momentum and magnetic dipole moment. Further, these external factors will also induce several new terms of spin angular momentum and magnetic dipole moment, for the quarks and gluons within the proton, in case the external electromagnetic fields and torque derivative are capable of impacting the inside of proton.

Within the proton, the quarks and gluons themselves possess some electric charges, and they are in constant motion all the while. Because there are several spatial and diversiform structural configurations, among some quarks and gluons, the influences of these external factors inside a proton will be different to that outside a proton. As a result, new increasing components of spin angular momentum and magnetic dipole moment, induced by these external factors, inside a proton must be different to that outside the proton.

The intervention of these external factors increases not only new components of spin magnetic moment in the magnetic dipole moment, but also new components of other terms in the octonion angular momentum. Therefore the octonion angular momentum, after the intervention of these external factors, should be different to that before the intervention of these external factors. In other words, applying the electromagnetic measuring method to measure the electromagnetic systems must alter the orbital angular momentum and electric/magnetic dipole moment and so forth for the octonion angular momentum.

\subsection{Decomposition}

In the complex octonion space, according to the octonion angular momentum, the proton is seized of the orbital angular momentum $\textbf{L}_g$ and magnetic dipole moment $\textbf{L}_e$ and so on. In normal circumstances, either of the orbital angular momentum $\textbf{L}_g$ and magnetic dipole moment $\textbf{L}_e$ may contain several components. There is no exact proportional relation between the orbital angular momentum $\textbf{L}_g$ and magnetic dipole moment $\textbf{L}_e$ , in general. After neglecting a few tiny terms, it is possible to possess an exact proportional relation between the major constituent of the orbital angular momentum $\textbf{L}_g$ and that of the magnetic dipole moment $\textbf{L}_e$ . Therefore, we should measure the orbital angular momentum and magnetic dipole moment for the proton respectively and independently, considering the interferences coming from some other physical quantities.

For the octonion angular momentum of the proton, neither of the orbital angular momentum $\textbf{L}_g$ and magnetic dipole moment $\textbf{L}_e$ is an invariant. However, the norm of octonion angular momentum for the proton is an invariant. In other words, the norm of octonion angular momentum, for the sub-particles within a proton, should be equal to that outside the proton. The scalar part of the octonion angular momentum is also an invariant.

Before the measurement, the octonion angular momentum for the proton is $\mathbb{L}_{(existing)}$ . In course of the measurement, the external factors will exert an influence on the proton, inducing a new component, $\mathbb{L}_{(new)}$ , of octonion angular momentum. Finally, the octonion angular momentum becomes, $\mathbb{L} = \mathbb{L}_{(existing)} + \mathbb{L}_{(new)}$. According to the definition, the octonion angular momentum is separated into, $\mathbb{L} = \mathbb{L}_g + k_{eg} \mathbb{L}_e$ . The quaternion angular momentum is, $\mathbb{L}_g = L_{g0} + i \textbf{L}_g^i + \textbf{L}_g$ , and the $S$-quaternion electromagnetic moment is, $\mathbb{L}_e = \textbf{L}_{e0} + i \textbf{L}_e^i + \textbf{L}_e$ . Especially, the orbital angular momentum is increased from the existing component $\textbf{L}_{g(existing)}$ to, $\textbf{L}_g = \textbf{L}_{g(existing)} + \textbf{L}_{g(new)}$ , with $\textbf{L}_{g(new)}$ being the induced new component of orbital angular momentum. And the magnetic dipole moment is increased from the existing component $\textbf{L}_{e(existing)}$ to, $\textbf{L}_e = \textbf{L}_{e(existing)} + \textbf{L}_{e(new)}$ , with $\textbf{L}_{e(new)}$ being the new induced component of magnetic dipole moment.

\subsection{Quarks and gluons}

In the vacuum without any external field, the proton is in a state of free movement, and is not subject to any interference of external factors. In this case, the octonion angular momentum, of the quarks and gluons within the proton, should be equal to that outside the proton, $\mathbb{L}_{(vacuum)}$ . However, in case there are some external factors exert an influence on the proton, the octonion angular momentum, of the quarks and gluons within a proton, should be still equal to that outside the proton, $\mathbb{L}_{(field)}$ . Apparently, the octonion angular momentum $\mathbb{L}_{(vacuum)}$ may not be equal to the octonion angular momentum $\mathbb{L}_{(field)}$ .

When the external electromagnetic fields exert an influence on the proton, it may not be able to impact effectively the quarks and gluons within the proton. By reason of the existence of electromagnetic fields of quarks and gluons themselves within a proton, the influence of external electromagnetic fields inside the proton may be quite tiny, inducing merely a little new octonion angular momentum, including few new spin angular momentum and other terms. In other words, the new spin angular momentum of quarks and gluons may be smaller than that outside the proton, because the orbital angular momentum, $\textbf{L}_g$ , is not an invariant either, in the complex octonion space.

In contrast to the new spin angular momentum outside the proton, the new spin angular momentum of quarks and gluons may be quite tiny. Besides the influence of distribution configuration of the external electromagnetic fields, it may be related with the direct contribution of external electromagnetic fields. The contribution of the external electromagnetic fields to the quarks and gluons, within the proton, may enhance the existing spin angular momentum sometimes, otherwise may weaken it. And it may be similar to the situation that the external magnetic field impacts the existing magnetic field inside the metal conductors.

\subsection{Polarization}

In the complex octonion space, according to the precessional equilibrium equation, Eq.(21), the external magnetic field (or electric field, torque derivative and so on) must induce the extra angular velocity of precession and the extra component of angular momentum. These induced components of angular momentum may be seized of various quantum numbers, including the half quantum number and others. Undoubtedly the spin angular momentum (or component of angular momentum) with the half quantum number plays an important role in the quantum theory. And the contributions of the components of angular momentum with rest quantum numbers may be comparatively tiny, which may even be regarded as the sources of background noises occasionally.

Let us suppose that the orientations of orbital angular momentum, for the ions in an ion beam, are uniform distribution with respect to the azimuth angles. That is, the orientations of orbital angular momenta are isotropy, in a plane perpendicular to the ion beam. When one ion beam passes through the external magnetic field with the azimuth angle $0^\circ$ , it is able to induce an extra component of angular momentum with the azimuth angle $0^\circ$ , resulting in the ion beam to be deviated from the uniform distribution with respect to the azimuth angle to a certain extent. This is the ion beam $A_{(ion)}$. Similarly, in case another ion beam passes through the external magnetic field, with the azimuth angle $\alpha^\circ$ , it will induce another extra component of angular momentum with the azimuth angle $\alpha^\circ$ , and this is the ion beam $B_{(ion)}$.

Based on the following reasons, the polarization difference between the angular momentums of two ion beams, $A_{(ion)}$ and $B_{(ion)}$, is indistinctive. Firstly, according to the definition of octonion angular momentum, the orbital angular momentum involves in many independent arguments, including the radius vector, integrating function of field potential, linear momentum, and electric current and so on. The magnetic field is merely one of arguments, and is not strong enough to cause a distinct discrepancy, between the angular momentums of two ion beams, $A_{(ion)}$ and $B_{(ion)}$, even if $\alpha = 180^\circ$ . Secondly, after the ion beam passes through the effective range of magnetic field, the influence of external magnetic field clears away, while other influence factors will spread out the induced angular velocity of precession in the ion beam gradually, tending the ion beams, $A_{(ion)}$ and $B_{(ion)}$, to be the same to a certain extent. Thirdly, within the collision zone, if the circumjacent external magnetic fields remain the same, their interferences to two ion beams, $A_{(ion)}$ and $B_{(ion)}$, will be identical.

In the complex octonion space, for the proton, the orbital angular momentum $\textbf{L}_g$ and magnetic dipole moment $\textbf{L}_e$ both contain several arguments. In general, there is not a proportion relation between two terms, $\textbf{L}_g$ and $\textbf{L}_e$, they should be measured independently. It means that the existing studies, for the proton spin puzzle, may skip over several possible variables, which had never been considered before. At present, we are still only at the foothills of probing into the proton spin puzzle, and there are something remaining to be explored.

\section{Experiment proposal}

In the complex octonion space, it is quite tough to measure the precessional motion of one neutral particle, when there is only the gravitational field. Contrastively it may be much more feasible to validate a few precessional motions, for one charged particle in the gravitational and electromagnetic fields. According to the precessional equilibrium equation, Eq.(21), a charged particle is able to yield a few precessional angular velocities in the electromagnetic field, inducing some new components of the magnetic dipole moment, revealing several aspects of the physical property relevant to the spin magnetic moment and spin angular momentum.

1) In the magnetic field, a stationary charged particle will possess the precessional angular velocity. The stationary charge may take part in the construction of the physical quantity, $\textbf{W}_{e0}^i$ . Therefore the interacting term, $\textbf{W}_{e0}^i \circ \textbf{B}$ , is capable of producing one precessional angular velocity with $k = 3$, according to Eq.(22). On the other hand, the stationary charge will take part in constituting the physical quantity, $\textbf{L}_{e0}$ . Meanwhile the interacting torque, $\textbf{L}_{e0} \circ \textbf{B}$, may cause the next precessional angular velocity with $k = 3$ , according to Eq.(24). On the basis of the existing Stern-Gerlach experiment, it may be feasible to fulfill these experiments. The precondition of experiment is that the electric current must be equal to zero, eliminating the possible interference of the Lorentz force.

2) In the electric field, the electric current will own the precessional angular velocity, when its net electric charge is zero. The electric current can take part in the construction of the physical quantity, $\textbf{W}_e$ . According to the precessional equilibrium equation, $\textbf{N}_g = 0$, the interacting term, $\textbf{E} \times \textbf{W}_e$ , will induce the precessional angular velocity. That is, the electric current may generate the precession motion in the electric field. On the basis of the magnetic dipole moment or the planar circular motion in the uniform magnetic field, it may be possible to validate the precessional motion of electric current in the electric field. The precondition is that there is no net electric charge of the electric current, obviating the latent disturbance of the Coulomb force.

3) In the Spintronics of the condensed matter physics, the electromagnetic strength inside the matters is generally hundreds or a thousand times stronger than that of the external fields in the lab. As a result, the external electromagnetic fields are incapable of discretionarily penetrating through the matters up to now, attempting to impact availably the spin angular momentum inside the matters neither. Presently, the scholars have to exploit some other physical properties of matters, to get hold of the spin-polarized electron beam. In the near future, in case it may be able to heighten dramatically the external electromagnetic strength (such as, thousands times), the scholars may alter directly the existing electron spin angular momentum or induce the new electron spin angular momentum in this super-strong external electromagnetic field, allowing us to generate the spin-polarized electron beam more easily than ever before. The application of this physical property is able to design an `amplifier' for the spin-polarized electron beam, achieving the function of repeater, extending the spin-diffusion length of the spin-polarized electron beam. Apparently, this ¡°amplifier¡± for the spin-polarized electron beam is similar to the all-optical amplifier in the optical communication.

The validation of the above experiment proposals will be of benefit to investigate the further features of spin angular momentum, decomposing correctly the octonion angular momentum of the proton.

\section{Discussions and conclusions}

In the complex octonion space $\mathbb{O}$ , the wave function, Eq.(36) , is relevant to the octonion torque. The possible contribution of the octonion torque to the composite operator may come from the electromagnetic and gravitational fields. Obviously each term of the octonion torque may have an influence on the precessional motion.

According to Eq.(9), there is only the torque between the term $\textbf{L}_{e0}$ with the external magnetic flux density, in the complex octonion space $\mathbb{O}$ . In the quantum theory described with the complex octonion, two terms, $\textbf{L}_{e0}$ and $\textbf{W}_{e0}^i$, will compel the charged particle to experience the Larmor precession, resulting in the precessional motion around the direction, $\textbf{\emph{I}}_0 \circ \textbf{B}$ , which is in the space $\mathbb{H}_g$ . That is, the two terms, $\textbf{L}_{e0}$ and $\textbf{W}_{e0}^i$ , determine the angular velocity of precession and the precessional direction. This inference is able to explain some problems. Such as, why does the precessional motion often revolve around the external magnetic flux density, for the charged particle? Who does provide the angular velocity of precession, and where does it originate from?

The external magnetic field is capable of interacting with two terms, $\textbf{L}_{e0}$ and $\textbf{W}_{e0}^i$ , to induce the magnetic dipole moment to experience the precessional motion, varying the direction of magnetic dipole moment with respect to the external magnetic flux density, altering the energy (and the torque) between the magnetic dipole moment with the external magnetic field. From the precessional equilibrium equation, $\textbf{N}_g = 0$ , some other terms of $\mathbb{W}$ will also exert an influence on the precession of charged particle, resulting in certain complicated observational results, as shown in the Stern-Gerlach and Zeeman experiments.

In the complex octonion space, according to the precessional equilibrium equation, $\textbf{N}_g = 0$, a part of spin angular momentums may originate from the precessional motions, induced by the magnetic field and term $\textbf{W}_{e0}^i$ and so forth, possessing the half quantum number. Essentially, the spin angular momentum is in the possession of the orbital angular momentum $\textbf{L}_g$ , meanwhile the spin magnetic moment belongs to the magnetic dipole moment $\textbf{L}_e$ . The octonion angular momentum, Eq.(5), can be separated into the orbital angular momentum, magnetic dipole moment, and electric dipole moment and so on. It reveals that the decomposition of the octonion angular momentum of a proton will be more delicate than each of four existing decompositions (Jaffe-Manohar, Ji, Chen, and Wakamatsu) for the proton angular momentum. Especially, the orbital angular momentum $\textbf{L}_g$ and magnetic dipole moment $\textbf{L}_e$ are independent of each other. In the proton spin puzzle, the orbital angular momentum $\textbf{L}_g$ and magnetic dipole moment $\textbf{L}_e$ must be measured and calculated respectively and independently.

It should be noted that the paper discussed only some simple cases about the precessional motions, proton spin puzzle, spin magnetic moment, and certain invariants, in the electromagnetic and gravitational fields described with the complex octonion. But it clearly states that the complex octonion is capable of describing availably the octonion angular momentum of a proton, constructing the physical model of spin angular momentum, emphasizing the independence between the magnetic dipole moment $\textbf{L}_e$ and orbital angular momentum $\textbf{L}_g$ . In the following study, it is going to probe into the influence of diversiform magnetic moment terms on several precessional motions, exploring the influence of the components of octonion torque on the wave equations, investigating some invariants relevant to the proton spin puzzle.

\section*{Acknowledgements}
The author is indebted to the anonymous referees for their constructive comments on the previous manuscript. This project was supported partially by the National Natural Science Foundation of China under grant number 60677039.

\appendix

\section{Exponential form}

The complex quaternion and octonion own a few properties, enabling the octonion quantities to be expressed as various exponential forms. Further the latter can be transformed into diverse wave functions. These features are beneficial to apply the complex octonion to express the quantum property of particles.

In the complex quaternion space $\mathbb{H}_g$ , a real quaternion, $\mathbb{Y}_g = y_0 + \Sigma \textbf{\emph{i}}_k y_k$, can be written as the exponential form,
\begin{eqnarray}
\mathbb{Y}_g = y_g exp ( \textbf{\emph{i}}_g \alpha_g )  ~,
\end{eqnarray}
where $\textbf{\emph{i}}_g = \Sigma \textbf{\emph{i}}_k y_k / ( \Sigma y_k^2 )^{1/2}$ . $y_g = ( \Sigma y_j^2 )^{1/2}$ . $cos \alpha_g = y_0 / y_g$ , $sin \alpha_g = ( \Sigma y_k^2 )^{1/2} / y_g$. $\textbf{\emph{i}}_g$ is one unit vector, with $\textbf{\emph{i}}_g^2 = -1$. $\alpha_g$ and $y_j$ are all real.

In the complex octonion space $\mathbb{O}$ , a real octonion, $\mathbb{Y} = \mathbb{Y}_g + k_y \mathbb{Y}_e$ , is able to be written as the exponential form,
\begin{eqnarray}
\mathbb{Y} = y_g exp ( \textbf{\emph{i}}_g \alpha_g ) + k_y \{ y_e exp ( \textbf{\emph{i}}_e \alpha_e ) \} \circ \textbf{\emph{I}}_0   ~ ,
\end{eqnarray}
where $\mathbb{Y}_e = ( Y_0 + \Sigma \textbf{\emph{i}}_k Y_k ) \circ \textbf{\emph{I}}_0$ , and it can be rewritten as the exponential form, $\mathbb{Y}_e = \{ y_e exp ( \textbf{\emph{i}}_e \alpha_e ) \} \circ \textbf{\emph{I}}_0$ . $\textbf{\emph{i}}_e = \Sigma \textbf{\emph{i}}_k Y_k / ( \Sigma Y_k^2 )^{1/2}$ . $y_e = ( \Sigma Y_j^2 )^{1/2}$. $cos \alpha_e = Y_0 / y_e$, and $sin \alpha_e = ( \Sigma Y_k^2 )^{1/2} / y_e$ . $\textbf{\emph{i}}_e$ is one unit vector, with $\textbf{\emph{i}}_e^2 = -1$. $k_y$ is one coefficient. $\alpha_e$ and $Y_j$ are all real.

In the paper, the complex octonion quantities possess a part of complex coordinate values, including the octonion angular momentum, torque, and force and so on. In the complex octonion space, one complex octonion physical quantity is,
\begin{eqnarray}
\mathbb{Z} = \mathbb{Z}_g + k_{eg} \mathbb{Z}_e  ~ ,
\end{eqnarray}
with
\begin{eqnarray}
&& \mathbb{Z}_g = i Z_{g0}^i + Z_{g0} + i \textbf{Z}_g^i + \textbf{Z}_g  ~,
\\
&& \mathbb{Z}_e = i \textbf{Z}_{e0}^i + \textbf{Z}_{e0} + i \textbf{Z}_e^i + \textbf{Z}_e  ~,
\end{eqnarray}
where $\textbf{Z}_g = \Sigma Z_{gk} \textbf{\emph{i}}_k$ . $\textbf{Z}_g^i = \Sigma Z^i_{gk} \textbf{\emph{i}}_k$ . $\textbf{Z}_{e0}^i = Z_{e0}^i \textbf{\emph{I}}_0$ . $\textbf{Z}_{e0} = Z_{e0} \textbf{\emph{I}}_0$ . $\textbf{Z}_e = \Sigma Z_{ek} \textbf{\emph{I}}_k$. $\textbf{Z}_e^i = \Sigma Z^i_{ek} \textbf{\emph{I}}_k$ . $Z_{gj}$ , $Z^i_{gj}$, $Z_{ej}$ , and $Z^i_{ej}$ are all real.

For the complex octonion physical quantity, $\mathbb{Z} = \mathbb{Z}_g + k_{eg} \mathbb{Z}_e$ , it can be rewritten as the exponential form,
\begin{eqnarray}
\mathbb{Z} = && { Z_1 exp ( i \alpha_1 ) + \textbf{Z}_2 exp ( i \alpha_2 ) }
\nonumber
\\
&&~~
+ k_{eg} \{ Z_3 exp ( i \alpha_3 ) + \textbf{Z}_4 exp ( i \alpha_4 ) \} \circ \textbf{\emph{I}}_0  ~ ,
\end{eqnarray}
where $Z_1 exp ( i \alpha_1 ) = i Z_{g0}^i + Z_{g0}$ ; $\textbf{Z}_2 exp ( i \alpha_2 ) = i \textbf{Z}_g^i + \textbf{Z}_g$; $ \{ Z_3 exp ( i \alpha_3 ) \} \circ \textbf{\emph{I}}_0 = i \textbf{Z}_{e0}^i + \textbf{Z}_{e0}$; $ \{ \textbf{Z}_4 exp ( i \alpha_4 ) \} \circ \textbf{\emph{I}}_0 = i \textbf{Z}_e^i + \textbf{Z}_e$. $\alpha_1$ , $\alpha_2$ , $\alpha_3$ , and $\alpha_4$ are all real. $\textbf{Z}_2$ and $\textbf{Z}_4$ both are real vectors in the space $\mathbb{H}_g$ . $Z_1$ and $Z_3$ are all real.

Also the above can be expressed as,
\begin{eqnarray}
\mathbb{Z} = && { i Z_5 exp ( \textbf{\emph{i}}_5 \alpha_5 ) + Z_6 exp ( \textbf{\emph{i}}_6 \alpha_6 ) }
\nonumber
\\
&&~~
+ k_{eg} \{ i Z_7 exp ( \textbf{\emph{i}}_7 \alpha_7 ) + Z_8 exp ( \textbf{\emph{i}}_8 \alpha_8 ) \} \circ \textbf{\emph{I}}_0 ~ ,
\end{eqnarray}
where $Z_5 exp ( \textbf{\emph{i}}_5 \alpha_5 ) = Z_{g0}^i + \textbf{Z}_g^i$ ; $Z_6 exp ( \textbf{\emph{i}}_6 \alpha_6 ) = Z_{g0} + \textbf{Z}_g$; $ \{ Z_7 exp ( \textbf{\emph{i}}_7 \alpha_7 ) \} \circ \textbf{\emph{I}}_0 = \textbf{Z}_{e0}^i + \textbf{Z}_e^i$; $ \{ Z_8 exp ( \textbf{\emph{i}}_8 \alpha_8 ) \} \circ \textbf{\emph{I}}_0  = \textbf{Z}_{e0} + \textbf{Z}_e $. $\alpha_5$ , $\alpha_6$ , $\alpha_7$ , and $\alpha_8$ are all real. $\textbf{\emph{i}}_5$, $\textbf{\emph{i}}_6$, $\textbf{\emph{i}}_7$, and $\textbf{\emph{i}}_8$ are unit vectors in $\mathbb{H}_g$ . $Z_5$ , $Z_6$ , $Z_7$ , and $Z_8$ are all real.

\section{Octonion wave function}

In the complex octonion space $\mathbb{O}$ , one wave equation associated with the octonion angular momentum, $\mathbb{L}$ , may be defined as,
\begin{eqnarray}
\Psi_L = \mathbb{L} / \hbar  ~ ,
\end{eqnarray}
where $\Psi_L$ is one dimensionless quantity, while the component, $\Psi_{Lg} = \mathbb{L}_g / \hbar $ , is in the space $\mathbb{H}_g$ . $(2 \pi \hbar)$ is the Planck constant.

According to the definition of octonion angular momentum, there are, $L_{g0}^i = 0$ and $\textbf{L}_{e0}^i = 0$ . It means that, there is, $\alpha_5^\prime = \pi / 2$ , for $L_9 exp ( \textbf{\emph{i}}_9 \alpha_5^\prime ) = \textbf{L}_g^i / \hbar $ ; and there is, $\alpha_7^\prime = \pi / 2$ , for $\{ L_9^\prime exp ( \textbf{\emph{i}}_9^\prime \alpha_7^\prime ) \} \circ \textbf{\emph{I}}_0 = \textbf{L}_e^i / \hbar $ . So the wave function component, $\Psi_{Lg}$ , can be rewritten as,
\begin{eqnarray}
\Psi_{Lg} = && i \textbf{L}_g^i / \hbar + L_{g0} / \hbar + \textbf{L}_g / \hbar
\nonumber
\\
~~~~~
= && i L_9 exp ( \textbf{\emph{i}}_9 \pi / 2) + L_q exp ( \textbf{\emph{i}}_q \alpha_q ) ~ ,
\end{eqnarray}
with
\begin{eqnarray}
\alpha_q = arc cos \{ [ ( P_{g0} \textbf{\emph{i}}_0 ) \cdot ( r_0 \textbf{\emph{i}}_0 )
+ ( \Sigma P_{gk} \textbf{\emph{i}}_k ) \cdot ( \Sigma r_k \textbf{\emph{i}}_k ) ] / ( \hbar L_q ) \}  ~,
\end{eqnarray}
where $L_q exp ( \textbf{\emph{i}}_q \alpha_q ) = ( L_{g0} + \textbf{L}_g ) / \hbar $ . $L_9$ , $L_9^\prime$ , $L_q$ , and $\alpha_q$ are all real. $\textbf{\emph{i}}_9$ , $\textbf{\emph{i}}_9^\prime$, and $\textbf{\emph{i}}_q$ are unit vectors in $\mathbb{H}_g$ .

If there are, $\mathbb{R}_e = 0$ , $\mathbb{X} = 0$ , $ \Sigma ( L_{gj}^i )^2  \ll \Sigma L_{gj}^2 $ , and $\mid L_{g0} \mid / ( \Sigma L_{gj}^2 )^{1/2} \ll 1$, the wave function $ \Psi_{Lg} $ will be approximately reduced into the wave function component, $\Psi_{Lg}^\prime = L_q exp ( \textbf{\emph{i}}_q \alpha_q )$ . The latter can be reduced to,
\begin{eqnarray}
\Psi_{Lg}^\prime \approx  L_q \textbf{ \emph{i}}_q \circ exp ( - \textbf{\emph{i}}_q \alpha_q^\prime )  ~,
\end{eqnarray}
further, making use of the transformation, $\Psi = - \textbf{\emph{i}}_q \circ \Psi_{Lg}^\prime$ , we have,
\begin{eqnarray}
\Psi =  L_q exp ( - \textbf{\emph{i}}_q \alpha_q^\prime )  ~,
\end{eqnarray}
with
\begin{eqnarray}
\alpha_q^\prime = [ ( P_{g0} \textbf{\emph{i}}_0 ) \cdot ( r_0 \textbf{\emph{i}}_0 ) + ( \Sigma P_{gk} \textbf{\emph{i}}_k ) \cdot ( \Sigma r_k \textbf{\emph{i}}_k ) ] / ( \hbar L_q ) ~.
\end{eqnarray}

In the complex octonion space $\mathbb{O}$ , when the direction of unit vector $\textbf{\emph{i}}_q$ is incapable of playing a major role in the wave function, this unit vector can be replaced by the imaginary unit, $i$ . As a result, the above will be degenerated into one familiar wave function, $ \Psi_{(4)} = exp \{ - i ( P_{g0} r_0 - \Sigma P_{gk} r_k ) / \hbar \} $, with $L_q = 1$, in the Quantum Mechanics. In other words, making use of the transformation, $\Psi = - \textbf{\emph{i}}_q \circ \Psi_{Lg}^\prime$ , and the restrictive condition, $\mid L_{g0} \mid / ( \Sigma L_{gj}^2 )^{1/2} \ll 1$, it is able to obtain the familiar wave function. Further, it is able to achieve diverse wave functions by means of various transformations. For example, one may substitute a vector $\textbf{z}$ for the unit vector $\textbf{\emph{i}}_q$ in the transformation, $\Psi = - \textbf{\emph{i}}_q \circ \Psi_{Lg}^\prime$, to yield a new wave function, $\Psi^{\prime\prime} = - \textbf{z} \circ \Psi_{Lg}^\prime$. Similarly, some wave functions in the quantum theory described with the complex octonion can be degenerated into that in the Quantum Mechanics, including the familiar wave function, $ \Psi_{(4)}^\prime = A_{(4)} exp \{ - i ( P_{g0} r_0 - \Sigma P_{gk} r_k ) / \hbar \} $, with $A_{(4)}$ being a scalar.

By contrast with the wave property derived from the mechanical vibration, the wave property of wave function, $\Psi$, is much more complicated and diverse obviously. Especially, the wave function, $ \Psi_L = \mathbb{L} / \hbar $, can be applied to explore the quantum property with respect to the orbital angular momentum, electric dipole moment, and magnetic dipole moment and other terms, because the octonion angular momentum, $\mathbb{L}$, consists of these physical quantities.

\section{Wave equation}

It is able to combine the octonion torque, $\mathbb{W}$ , and quaternion operator together to create a new composite operator, $\{ i \mathbb{W}^\star / ( \hbar v_0 ) + \lozenge \}$ . When this new composite operator acts on a few physical quantities in Table 3, it is able to infer some equations of the quantum theory in the complex octonion space, including the wave function and wave equation.

In the definition of octonion torque, substituting the operator, $\{ i \mathbb{W}^\star / ( \hbar v_0 ) + \lozenge \}$, and $( \mathbb{Z}_L \circ \mathbb{L} )$ , for the operator, $( i \mathbb{F} / v_0 + \lozenge )$, and $\mathbb{L}$ respectively, will yield one wave function relevant to the octonion torque as follows,
\begin{eqnarray}
\Psi_{WL} = - v_0 \{ i \mathbb{W}^\star / ( \hbar v_0 ) + \lozenge \} \circ ( \mathbb{Z}_L \circ \mathbb{L} )  ~ ,
\end{eqnarray}
or
\begin{eqnarray}
\Psi_{WL} = - ( i \mathbb{W}^\star + \hbar v_0 \lozenge ) \circ \Psi_{ZL}  ~ ,
\end{eqnarray}
where $ \Psi_{ZL} = ( \mathbb{Z}_L \circ \mathbb{L} ) / \hbar $ . The octonion torque $\mathbb{W}$ and octonion angular momentum $\mathbb{L}$ both comprise the contributions coming from the electromagnetic and gravitational fields.

When a particle situates in the external electromagnetic and gravitational fields, Eq.(C.2) is one wave function relevant to the octonion torque, for the particle with the nonzero rest mass. Especially, in case $\Psi_{WL} = 0$, one wave equation relevant to the octonion torque will be derived from Eq.(C.2). Further this wave equation can be degenerated into the Dirac wave equation. When there is no external electromagnetic and gravitational fields, the equation, $\Psi_{WL} = 0$ , will be simplified to the wave equation, for the particle with the nonzero rest mass in a free motion state.

\section{Dirac wave equation}

In the complex octonion space $\mathbb{O}$ , it is capable of deducing the Dirac wave equation, from the wave equation, $\Psi_{WL} = 0$ , relevant to the octonion torque, under a certain approximate conditions. In the quantum theory described with the complex octonion, the electromagnetic potential and gravitational potential both make a contribution towards the Dirac wave equation to a certain extent.

In case there is only the electromagnetic potential, the octonion torque is able to be reduced to,
\begin{eqnarray}
\mathbb{W} \approx i W_{g0}^i + \textbf{W}_g  ~ ,
\end{eqnarray}
with
\begin{eqnarray}
&& W_{g0}^i \approx  k_p P_{g0} v_0 + k_{eg}^2 ( - \textbf{A}_{e0} \circ \textbf{P}_{e0} ) ~ ,
\\
&& \textbf{W}_g \approx k_{eg}^2 ( \textbf{A}_e \circ \textbf{P}_{e0} )   ~ ,
\end{eqnarray}
where $P_{g0} = m^\prime v_0$ . $\textbf{P}_{e0} = ( \mu_e / \mu_g ) q V_0 \textbf{\emph{I}}_0 $ . $\mathbb{V}_g = v_0 \partial_0 \mathbb{R}_g$ . $\textbf{V}_0$ is the scalar-like part of $S$-quaternion velocity, $\mathbb{V}_e = v_0 \partial_0 \mathbb{R}_e $ , with $\textbf{V}_0 = V_0 \textbf{\emph{I}}_0$ . In general, $V_0 / v_0 \approx 1 $.

In case $\Psi_{WL} = 0$, substituting the above in Eq.(C.2) will deduce the Dirac wave equation described with the complex octonion,
\begin{eqnarray}
( i \hbar v_0 \partial_0 + W_{g0}^i  + \hbar v_0 \nabla + i \textbf{W}_g ) \circ \Psi_{ZL} = 0  ~ ,
\end{eqnarray}
for the charged particle in the electromagnetic field.

Further the above can be expanded into,
\begin{eqnarray}
&& \{ i \hbar v_0 \partial_0 - k_{eg}^2 ( \textbf{A}_{e0} \circ \textbf{P}_{e0} ) + \hbar v_0 \nabla
\nonumber
\\
&&~~
 +  i k_{eg}^2 ( \textbf{A}_e \circ \textbf{P}_{e0} ) + k_p m^\prime v_0^2  \} \circ \Psi_{ZL} = 0  ~ ,
\end{eqnarray}
in the complex octonion space $\mathbb{O}$ .

When there are the electromagnetic and gravitational potentials simultaneously, some terms in Eq.(D.1) can be extended approximately into,
\begin{eqnarray}
&& W_{g0}^i \approx  k_p P_{g0} v_0 + k_{eg}^2 ( - \textbf{A}_{e0} \circ \textbf{P}_{e0} ) - A_{g0} P_{g0}  ~ ,
\\
&& \textbf{W}_g \approx k_{eg}^2 ( \textbf{A}_e \circ \textbf{P}_{e0} ) + P_{g0} \textbf{A}_g   ~ .
\end{eqnarray}

As a result, Eq.(D.1) can be expanded into,
\begin{eqnarray}
\mathbb{W}  \approx && i k_p m^\prime v_0^2 - i ( k_{eg}^2 \textbf{A}_{e0} \circ \textbf{P}_{e0} +  A_{g0} P_{g0} )
\nonumber
\\
&&~~
+ k_{eg}^2 \textbf{A}_e \circ \textbf{P}_{e0} +  P_{g0} \textbf{A}_g    ~ ,
\end{eqnarray}
where either of terms, $\mathbb{A}_g$ and $P_{g0}$ , contains the contribution of gravitational fields. And $\mathbb{A}_e$ and $\textbf{P}_{e0}$ both consist of the contribution of electromagnetic fields.

In case $\Psi_{WL} = 0$, substituting the above in the wave function, Eq.(C.2), which is relevant to the octonion torque, will infer the Dirac wave equation described with the complex octonion,
\begin{eqnarray}
&& \{  i \hbar v_0 \partial_0 - ( k_{eg}^2 \textbf{A}_{e0} \circ \textbf{P}_{e0} +  A_{g0} P_{g0} ) +  \hbar v_0 \nabla
\nonumber
\\
&&~~
 +  i ( k_{eg}^2 \textbf{A}_e \circ \textbf{P}_{e0} +  P_{g0} \textbf{A}_g  )  + k_p m^\prime v_0^2  \} \circ \Psi_{ZL} = 0  ~ ,
\end{eqnarray}
for the charged particle in the electromagnetic and gravitational fields.

The above states that the gravitational potential and electromagnetic potential both make a contribution to the wave equation relevant to the octonion torque, when there are the gravitational and electromagnetic fields.

\section{Schr\"{o}dinger wave equation}

Based on the wave function, Eq.(C.2), relevant to the octonion torque, it is able to deduce not only the Dirac wave equation but also the Schr\"{o}dinger wave equation. Meanwhile each of the gravitational field and electromagnetic field will make a contribution to the Schr\"{o}dinger wave equation to a certain extent.

In order to derive the Schr\"{o}dinger wave equation from the Dirac wave equation, it is necessary to introduce one appropriate operator, $\lozenge_{DS}$ . When this operator acts on the wave function, Eq.(C.2), from the left side, we will obtain one new wave function,
\begin{eqnarray}
\Psi_{WL}^\prime = \lozenge_{DS} \circ \Psi_{WL}  ~ ,
\end{eqnarray}
or
\begin{eqnarray}
\Psi_{WL}^\prime =  - v_0 \{  i \mathbb{W}_{DS} / ( \hbar v_0 ) + \lozenge \} \circ \{ [ i \mathbb{W}^\star / ( \hbar v_0 ) + \lozenge ] \circ ( \mathbb{Z}_L \circ \mathbb{L} ) \} ~ ,
\end{eqnarray}
where $ \lozenge_{DS} = i \mathbb{W}_{DS} / ( \hbar v_0 )+ \lozenge $ , with $ \mathbb{W}_{DS} = \mathbb{W}^\star + i ( 2 k_p m^\prime v_0^2 ) $ .

From the above, it is able to deduce the Schr\"{o}dinger wave equation, described with the complex octonion, for the charged particle, when $\Psi_{WL}^\prime = 0$.

The octonion possesses the nonassociative and noncommutative properties, but it still satisfies the weak associative property. And the quaternion has the associative and noncommutative properties. For two octonions, $\mathbb{Z}$ and $\mathbb{Y}$ , the weak associative property between them is,
\begin{eqnarray}
\mathbb{Z} \circ ( \mathbb{Z} \circ \mathbb{Y} ) = ( \mathbb{Z} \circ \mathbb{Z} ) \circ \mathbb{Y}  ~ .
\end{eqnarray}

The weak associative property of octonions and the associative property of quaternions will be great of benefit to simplify some wave functions. As a result, the complex octonion wave function can be written as,
\begin{eqnarray}
\Psi_{WL}^\prime = && - v_0 \{ [ i \mathbb{W}^\star / ( \hbar v_0 ) + \lozenge ] \circ [ i \mathbb{W}^\star / ( \hbar v_0 ) + \lozenge ] \} \circ ( \mathbb{Z}_L \circ \mathbb{L} )
\nonumber
\\
&&~~~~
- v_0 [ - 2 k_p m^\prime v_0^2 / ( \hbar v_0 ) ] \circ [ i \mathbb{W}^\star / ( \hbar v_0 ) + \lozenge ] \circ ( \mathbb{Z}_L \circ \mathbb{L} )  ~ .
\end{eqnarray}

In the first term of the right side, the product among components of the operator can be expanded into,
\begin{eqnarray}
&&~~ \{ i \mathbb{W}^\star / ( \hbar v_0 ) + \lozenge \} \circ \{ i \mathbb{W}^\star / ( \hbar v_0 ) + \lozenge \}
\nonumber
\\
= && \{ [ W_{g0}^i / ( \hbar v_0 ) + i \partial_0 ] + ( \textbf{W}_g^i + k_{eg}^\prime \textbf{W}_{e0} + k_{eg}^\prime \textbf{W}_e ) / ( \hbar v_0 )
\nonumber
\\
&& ~~~~~~
+ [ i \textbf{W}_g / ( \hbar v_0 ) + \nabla ] + i ( W_{g0} - k_{eg}^\prime \textbf{W}_{e0}^i - k_{eg}^\prime \textbf{W}_e^i ) / ( \hbar v_0 ) \}
\nonumber
\\
&& ~~ \circ \{ [ W_{g0}^i / ( \hbar v_0 ) + i \partial_0 ] + ( \textbf{W}_g^i + k_{eg}^\prime \textbf{W}_{e0} + k_{eg}^\prime \textbf{W}_e ) / ( \hbar v_0 )
\nonumber
\\
&& ~~~~~~
+ [ i \textbf{W}_g / ( \hbar v_0 ) + \nabla ] + i ( W_{g0} - k_{eg}^\prime \textbf{W}_{e0}^i - k_{eg}^\prime \textbf{W}_e^i ) / ( \hbar v_0 ) \}  ~ ,~~
\end{eqnarray}
where $k_{eg} = i k_{eg}^\prime$ , and $k_{eg}^\prime$ is real.

Neglecting some tiny terms in the above, and considering the contribution of following terms,
\begin{eqnarray}
&&~~~~ ( \nabla \times \textbf{W}_g + \partial_0 \textbf{W}_g^i ) / ( \hbar v_0 )
\nonumber
\\
\approx && k_{eg}^2 \nabla \times ( \textbf{B} \times \textbf{L}_e^i + \textbf{A}_e \circ \textbf{P}_{e0} ) / ( \hbar v_0 ) + k_{eg}^2 \partial_0 ( - \textbf{B} \circ \textbf{L}_{e0} ) / ( \hbar v_0 )
\nonumber
\\
&&
~~~ + \nabla \times ( \textbf{b} \times \textbf{L}_g^i + P_{g0} \textbf{A}_g ) / ( \hbar v_0 ) + \partial_0 ( - L_{g0} \textbf{b} ) / ( \hbar v_0 )
\nonumber
\\
\approx && k ( k_{eg}^2 \textbf{P}_{e0} \circ \textbf{B} + P_{g0} \textbf{b} ) / ( \hbar v_0 ) ~ ,
\end{eqnarray}
the octonion wave function, $ \Psi_{WL}^\prime $ , can be approximately reduced to,
\begin{eqnarray}
\Psi_{WL}^\prime \approx && - v_0 \{  [ E_w / ( \hbar v_0 ) + i \partial_0 ]^2 - [ k_p m^\prime v_0^2 / ( \hbar v_0 ) ]^2
\nonumber
\\
&&~~~~~~
+ [ i \textbf{W}_g / ( \hbar v_0 ) + \nabla ] \cdot [ i \textbf{W}_g / ( \hbar v_0 ) + \nabla ]
\nonumber
\\
&&~~~~~~
+ i k ( k_{eg}^2 \textbf{P}_{e0} \circ \textbf{B} + P_{g0} \textbf{b} ) / ( \hbar v_0 ) \}  \circ (\mathbb{Z}_L \circ \mathbb{L})   ~ ,
\end{eqnarray}
where the second term connects with the dimension of radius vector, $\textbf{r}$ , apparently. The energy part can be rewritten as, $W_{g0}^i = k_p m^\prime v_0^2 + E_w$ . The term $E_w$ includes $( \textbf{A}_0 \circ \textbf{P}_{e0} )$ , while the vector-like term $\textbf{W}_g$ contains $( \textbf{A}_e \circ \textbf{P}_{e0} )$.

On the basis of the above analysis, the Schr\"{o}dinger wave equation can be derived from the equation, $\Psi_{WL}^\prime = 0$ , and written as follows,
\begin{eqnarray}
&& \{ [ E_w / ( \hbar v_0 ) + i \partial_0 ]^2 - [ k_p m^\prime v_0^2 / ( \hbar v_0 ) ]^2
\nonumber
\\
&& ~~~~
+ [ i \textbf{W}_g / ( \hbar v_0 ) + \nabla ] \cdot [ i \textbf{W}_g / ( \hbar v_0 ) + \nabla ]
\nonumber
\\
&& ~~~~
+ i k ( k_{eg}^2 \textbf{P}_{e0} \circ \textbf{B} + P_{g0} \textbf{b} ) / ( \hbar v_0 )  \} \circ \Psi_{ZL} = 0 ~ .
\end{eqnarray}

Next we consider one wave function, $\Psi^{\prime\prime}$ , in the space $\mathbb{H}_g$ . In case the direction of unit vector, $\textbf{\emph{i}}_q$ , is incapable of playing a major role in the wave function $\Psi^{\prime\prime}$, this unit vector can be replaced by the imaginary unit, $i$ . As one component of the wave function $ \Psi_{ZL} $ , the wave function $ \Psi^{\prime\prime} $ can be chosen approximately as,
\begin{eqnarray}
\Psi^{\prime\prime} = \Psi (\textbf{r}) exp ( - i E_s t / \hbar ) ~ ,
\end{eqnarray}
where $E_s$ is the energy, and $\Psi (\textbf{r})$ is one complex vector function.

Substituting $\Psi^{\prime\prime}$ into Eq.(E.8) yields the wave equation,
\begin{eqnarray}
&& \{ - [ \textbf{W}_g - i ( \hbar v_0 ) \nabla ] \cdot [ \textbf{W}_g - i ( \hbar v_0 ) \nabla ]
\nonumber
\\
&& ~~~~
+ i k ( k_{eg}^2 \textbf{P}_{e0} \circ \textbf{B}  + P_{g0} \textbf{b} )
\nonumber
\\
&& ~~~~
- ( k_p m^\prime v_0^2 )^2 + ( E_w + k_p m^\prime v_0^2 + E^\prime )^2 \} \circ \Psi (\textbf{r}) = 0 ~ ,
\end{eqnarray}
where $ E_s = k_p m^\prime v_0^2 + E^\prime $, $ k_p m^\prime v_0^2 \gg E_w $, and  $ k_p m^\prime v_0^2 \gg E^\prime $.

Therefore the above can be approximately written as,
\begin{eqnarray}
&& \{ ( 2 k_p m^\prime v_0^2 ) ( E_w + E^\prime ) + i k ( k_{eg}^2 \textbf{P}_{e0} \circ \textbf{B} + P_{g0} \textbf{b} )
\nonumber
\\
&& ~~~~~
- [ \textbf{W}_g - i ( \hbar v_0 ) \nabla ] \cdot [ \textbf{W}_g - i ( \hbar v_0 ) \nabla ] \} \circ \Psi (\textbf{r}) = 0 ~ ,
\end{eqnarray}
and then there is,
\begin{eqnarray}
E^\prime \Psi (\textbf{r}) = && \{ [ \textbf{W}_g - i ( \hbar v_0 ) \nabla ] \cdot [ \textbf{W}_g - i ( \hbar v_0 ) \nabla ] /  ( 2 k_p m^\prime v_0^2 ) - E_w
\nonumber
\\
&& ~~~~
+ i k q V_0 \hbar \textbf{\emph{I}}_0 \circ \textbf{B} / ( 2 k_p m^\prime v_0 ) + i k \hbar \textbf{b} / ( 2 k_p )  \} \circ \Psi (\textbf{r}) ~ , ~~
\end{eqnarray}
where either of the last two terms is the vector term rather than the energy term, in the complex octonion space. The term, $ \textbf{L}_{e0(q)} \circ \textbf{B} $, is the torque between the magnetic flux density, $\textbf{B}$ , with the quantized term, $\textbf{L}_{e0(q)} = k q V_0 \hbar \textbf{\emph{I}}_0 / ( 2 k_p m^\prime v_0 ) $. Meanwhile the term, $ L_{g0(q)} \circ \textbf{b} $ , is the torque between the gravitational strength $\textbf{b}$ with the quantized term, $ L_{g0(q)} = k \hbar / ( 2 k_p ) $ .

\section{Half quantum number}

When the quaternion force equals to zero, that is, $ \mathbb{N}_g = 0 $ , there are the equations, $\textbf{N}_g = 0 $ and $\textbf{N}_g^i = 0 $. In terms of a charged particle, it is able to deduce the angular frequency of precession from the precessional equilibrium equation, $\textbf{N}_g = 0 $, meanwhile it is capable of inferring the angular frequency of revolution from the force equilibrium equation, $\textbf{N}_g^i = 0 $.

Let us consider one simple case. In the complex octonion space, one charged particle experiences the circular motion, within the uniform magnetic field. The force equilibrium equation, $\textbf{N}_g^i = 0$ , can be reduced into,
\begin{eqnarray}
k_{eg}^2 ( \textbf{B} \times \textbf{W}_e ) / v_0 - \partial_0 \textbf{W}_g = 0 ~,
\end{eqnarray}
where $ \textbf{W}_e \approx  k_p \textbf{P}_e v_0 $ . From the above, the angular frequency of revolution is, $ \omega_r = - B q V_0 / ( m v_0 ) $. Also it is the angular frequency of magnetic dipole moment $\textbf{L}_{e(r)}$ in the revolution motion.

In the uniform circular motion, there is no torque term, that is, $\textbf{W}_g^i = 0$. According to Eq.(22), the angular frequency of precession is,
\begin{eqnarray}
\omega_p = - B q V_0 / ( 2 m v_0 ) ~,
\end{eqnarray}
when $ k = 2$ , for the planar movement.

Obviously this precession will lead to one new component of magnetic dipole moment $\textbf{L}_{e(p)}$, with $ \omega_p $ being one half of $ \omega_r $ . As a result, the term, $\textbf{L}_{e(p)}$, can be considered as one major component of spin magnetic moment. And the energy gradient of the new energy term, $ \textbf{L}_{e(p)} \cdot \textbf{B} $, can be applied to explain the physical phenomenon in the Stern-Gerlach experiment, when $k = 2$ .

It means that the emergence of this new component of magnetic dipole moment $\textbf{L}_{e(p)}$ is accompanied by the existence of external magnetic field. Moreover it is still able to induce the new component of magnetic dipole moment for the charged particle, even if there is no magnetic dipole moment $\textbf{L}_{e(r)}$ nor the revolution motion, under the extreme condition that there is not any torque term.

When the magnetic dipole moment $\textbf{L}_{e(r)}$ is quantized and seized of a quantum number $l$ , the new component of magnetic dipole moment $\textbf{L}_{e(p)}$ will possess the half quantum number $(l/2)$ , due to the proportional relationship, $ \omega_p = \omega_r / 2 $, with $ k $ being 2. This simple case may be one common problem we often meet with.

\end{document}